\documentclass[preprint,notoc]{JHEP3}
\usepackage{epsfig}

\def\eslt{\not\!\!{E_T}}
\def\to{\rightarrow}

\def\bi{\begin{itemize}}
\def\ei{\end{itemize}}
\def\te{\tilde e}

\def\tu{\tilde u}

\def\tb{\tilde b}

\def\tst{\tilde t}
\def\ttau{\tilde \tau}
\def\tmu{\tilde \mu}
\def\tg{\tilde g}

\def\tell{\tilde\ell}
\def\tq{\tilde q}
\def\tB{\widetilde B}
\def\tw{\widetilde W}
\def\tz{\widetilde Z}
\def\alt{\stackrel{<}{\sim}}
\def\agt{\stackrel{>}{\sim}}
\def\be{\begin{equation}}  
\def\ee{\end{equation}}  

\title{Mixed Wino Dark Matter: Consequences for\\
Direct, Indirect and Collider Detection}

\author{Howard Baer, Azar Mustafayev, Eun-Kyung Park
and Stefano Profumo 
\\ Department of Physics, Florida State University Tallahassee, FL 32306, USA\\
E-mail: \email{baer@hep.fsu.edu},\email{mazar@hep.fsu.edu},
\email{epark@hep.fsu.edu},\email{profumo@hep.fsu.edu}}

\preprint{\vbox{\hbox{FSU-HEP-050530}}} 

\abstract{
In supersymmetric models with gravity-mediated SUSY breaking and
gaugino mass unification, the 
predicted relic abundance of neutralinos usually exceeds the strict
limits imposed by the WMAP collaboration. One way to obtain the correct
relic abundance is to abandon gaugino mass universality and allow
a mixed wino-bino lightest SUSY particle (LSP). The enhanced
annihilation and scattering cross sections of 
mixed wino dark matter (MWDM) compared to bino dark matter
lead to enhanced rates for direct dark matter detection, as well as
for indirect detection at neutrino telescopes and for detection of 
dark matter annihilation products in the galactic halo. 
For collider experiments, MWDM leads to a 
reduced but significant mass gap between the lightest neutralinos
so that $\tz_2$ two-body decay modes are usually closed. 
This means that dilepton mass
edges-- the starting point for cascade decay reconstruction at the CERN LHC--
should be accessible over almost all of parameter space.
Measurement of the $m_{\tz_2}-m_{\tz_1}$ mass gap at LHC plus 
various sparticle masses and cross sections as a function of beam polarization
at the International Linear Collider (ILC) would pinpoint MWDM as the dominant
component of dark matter in the universe.
}


\begin{document}

\section{Introduction}
\label{sec:intro}

In supersymmetric models of particle physics,
$R$-parity is often imposed to avoid too rapid proton decay
which can be induced by superpotential terms which violate
baryon and lepton number conservation. One of the byproducts of
$R$-parity conservation is that the lightest supersymmetric particle
is absolutely stable, making it a good candidate particle to make up 
the bulk of dark matter (DM) in the universe.
In gravity-mediated SUSY breaking models, dark matter candidate
particles include the lightest neutralino or the gravitino. Here we will
focus on the lightest neutralino $\tz_1$\cite{haim}; recent results on 
TeV scale gravitino dark matter can be found in Ref.~\cite{gravitino}.
The relic density of neutralinos in supersymmetric models can be 
straightforwardly calculated by solving the Boltzmann equation for the
neutralino number density\cite{griest}. 
The central part of the calculation is to
evaluate the thermally averaged neutralino annihilation and 
co-annihilation cross section times velocity. 
The computation requires evaluating many thousands of Feynman diagrams.
Several computer codes are now publicly\cite{rdcodes,isared} 
available which evaluate the
neutralino relic density $\Omega_{\tz_1}h^2$. 

The dark matter density of the universe has recently been inferred 
from the WMAP collaboration based on precision fits to 
anisotropies in the cosmic microwave background radiation\cite{wmap}.
The WMAP collaboration result for the relic density of cold dark matter (CDM)
is that
\begin{equation}
\Omega_{CDM}h^2=0.113\pm 0.009 .
\end{equation}
This result imposes a tight constraint on supersymmetric models
which contain a dark matter candidate\cite{wmapcon}. 

Many analyses have been 
recently performed in the context of the paradigm minimal
supergravity model\cite{msugra} (mSUGRA), where the parameter space is given by
$m_0,\ m_{1/2},\ A_0,\ \tan\beta$ and $sign(\mu )$. The mSUGRA model 
assumes the minimal supersymmetric model (MSSM) is valid between
the mass scales $Q=M_{GUT}$ and $Q=M_{SUSY}$. A common mass $m_0$ ($m_{1/2}$)
(($A_0$))
is assumed for all scalars (gauginos) 
((trilinear soft breaking parameters)) at $Q=M_{GUT}$, while the bilinear
soft term $B$ is traded for $\tan\beta$, the ratio of Higgs vevs, via the
requirement of radiative electroweak symmetry breaking (REWSB). REWSB
also determines the magnitude, but not the sign, of the superpotential
Higgs mass term $\mu$. Weak scale couplings and soft parameters
can be computed via renormalization group (RG) evolution from $Q=M_{GUT}$ to 
$Q=M_{weak}$. Once weak scale parameters are known, then sparticle masses 
and mixings may be computed, and the associated relic density of 
neutralinos can be determined.

In most of mSUGRA parameter space, the relic density $\Omega_{\tz_1}h^2$
turns out to be much larger than the WMAP value. Many analyses have found 
just several allowed regions of parameter space:
\begin{itemize}
\item The bulk region occurs at low values of $m_0$ and $m_{1/2}$\cite{bulk}. 
In this region, neutralino annihilation is enhanced by $t$-channel exchange of 
relatively light sleptons. The bulk region, featured prominently in many early 
analyses of the relic density, has been squeezed from below by the LEP2
bound on the chargino mass $m_{\tw_1}>103.5$ GeV, and from above by the 
tight bound from WMAP.
\item The stau co-annihilation region at low $m_0$ for almost any $m_{1/2}$
value where $m_{\ttau_1}\simeq m_{\tz_1}$, so that $\ttau_1-\tz_1$ and
$\ttau_1^+\ttau_1^-$ co-annihilation help to reduce the relic 
density\cite{stau}.
\item The hyperbolic branch/focus point (HB/FP) region at large $m_0\sim$ 
several TeV, where $\mu$ becomes small, and neutralinos efficiently 
annihilate via their higgsino components\cite{hb_fp}.
This is the case of mixed higgsino dark matter (MHDM).
\item The $A$-annihilation funnel occurs at large $\tan\beta$ values
when $2m_{\tz_1}\sim m_A$ and neutralinos can efficiently annihilate
through the broad $A$ and $H$ Higgs resonances\cite{Afunnel}.
\end{itemize}
In addition, a less prominent light Higgs $h$ annihilation corridor occurs at
low $m_{1/2}$\cite{drees_h} 
and a top squark co-annihilation region occurs at 
particular $A_0$ values when $m_{\tst_1}\simeq m_{\tz_1}$\cite{stop}.

Many analyses have also been performed for gravity-mediated SUSY breaking 
models with non-universal soft terms. Non-universality of 
SSB scalar masses can 1. pull various scalar masses to low values so that
``bulk'' annihilation via $t$-channel exchange of light 
scalars can occur\cite{nmh},
or 2. they can bring in new near degeneracies of various sparticles with the
$\tz_1$ so that new co-annihilation regions open up\cite{auto,nuhm,sp}, 
or they can 3.
bring the value of $m_A$ into accord with $2m_{\tz_1}$ so that 
funnel annihilation can occur\cite{ellis,nuhm}, 
or 4. they can pull the value of $\mu$
down so that higgsino annihilation can occur\cite{ellis,drees2,nuhm}. 
It is worthwhile noting that
all these general mechanisms for increasing the neutralino annihilation rate 
already occur in the mSUGRA model.
Moreover, in all these cases the lightest neutralino
is either bino-like, or a bino-higgsino mixture.

If non-universal gaugino masses are allowed, then a qualitatively new 
possibility arises that is not realized in the mSUGRA model: that of mixed
wino dark matter (MWDM). In this case, if the $SU(2)$ gaugino mass $M_2$ is 
sufficiently low compared to $U(1)_Y$ gaugino mass $M_1$, then the
$\tz_1$ can become increasingly wino-like. The $\tz_1-\tw_{1,2}-W$
coupling becomes large when $\tz_1$ becomes wino-like,
resulting in enhanced $\tz_1\tz_1\to W^+W^-$ annihilations. Moreover, coannihilations with the lightest chargino and with the next-to-lightest neutralino help to further suppress the LSP thermal relic abundance.

Non-universal gaugino masses can arise in supersymmetric models in a 
number of ways\cite{models}. 
\begin{itemize}
\item In supergravity GUT models, the gauge kinetic
function (GKF) $f_{AB}$ must transform as the symmetric product of two 
adjoints. 
In minimal supergravity, the GKF transforms as a singlet. In $SU(5)$
SUGRA-GUT models, it can also transform as a 24, 75 or 200 dimensional
representation\cite{anderson}, while in $SO(10)$ models it can transform as 
1, 54, 210 and 770 dimensional representations\cite{chamoun,nath}. 
Each of these
non-singlet cases leads to unique predictions for the ratios of
GUT scale gaugino masses. Furthermore, if the GKF transforms as a 
linear combination of these higher dimensional representations, then 
essentially arbitrary gaugino masses are allowed. 

\item Non-universal gaugino masses are endemic to heterotic superstring models
with orbifold compactification 
where SUSY breaking is dominated by the moduli fields\cite{ibanez}. 

\item Additionally, in extra-dimensional SUSY GUT models where SUSY breaking 
is communicated from the SUSY breaking brane to the visible brane via gaugino
mediation, various patterns of GUT scale gaugino masses can occur,
including the case of completely independent gaugino masses\cite{dermisek}.
\end{itemize}
In this report, we will adopt a phenomenological approach, and regard the
three MSSM gaugino masses as independent parameters, with the constraint 
that the neutralino relic density should match the WMAP measured value.

Much previous work has been done on evaluating the relic density in models with
gaugino mass non-universality.
In AMSB models\cite{amsb}, the
$\tz_1$ is almost pure wino, so that $\Omega_{\tz_1}h^2$ as predicted by
the Boltzmann equation is typically very low. 
Moroi and Randall\cite{moroi} proposed moduli
decay to wino-like neutralinos in the early universe to account for the
dark matter density. 
Already in 1991, Griest and Roszkowski had shown that a wide range of 
relic density values could be obtained by abandoning 
gaugino mass universality\cite{gr}.
Corsetti and Nath investigated dark matter relic density and detection rates 
in models with non-minimal $SU(5)$ GKF and 
also in O-II string models\cite{cor_nath}. Birkedal-Hanson and Nelson
showed that a GUT scale ratio $M_1/M_2\sim 1.5$ would bring the 
relic density into accord with the measured CDM density via MWDM, 
and also presented direct detection rates\cite{birkedal}.
Bertin, Nezri and Orloff showed variation of relic density and enhancement in
direct and indirect DM detection rates as non-universal gaugino masses 
were varied\cite{nezri}. Bottino {\it et al.} performed scans
over independent weak scale parameters to show variation in indirect 
DM detection rates, and noted that neutralinos as low as 6 GeV are
allowed\cite{bottino}. Belanger {\it et al.} presented relic density plots
in the $m_0\ vs. m_{1/2}$ plane for a variety of universal and 
non-universal gaugino mass scenarios, and showed that large swaths 
of parameter space open up when the $SU(3)$ gaugino mass $M_3$ becomes 
small\cite{belanger}. Mambrini and Munoz and also Cerdeno and Munoz 
showed direct and indirect detection rates for model with scalar and
gaugino mass non-universality\cite{munoz}. 
Auto {\it et al.}\cite{auto} used non-universal gaugino masses to reconcile 
the predicted relic density in models with Yukawa coupling unification 
with the WMAP result.
Masiero, Profumo and Ullio
exhibit the relic density and direct and indirect detection rates 
in split supersymmetry where $M_1$, $M_2$ and $\mu$ are taken as independent
weak scale parameters with ultra-heavy squarks and sleptons\cite{mpu}.
 
In this paper, we will adopt a model with GUT scale parameters
including universal scalar masses, but with
independent gaugino masses leading to MWDM. 
We will assume all gaugino masses to be of the same sign. The opposite sign 
situation leads to a distinct DM scenario and will be addressed
soon\cite{binodm}.
We will 
adjust the gaugino masses such that $\tz_1$ receives just enough of a 
wino component so that it makes up the entire CDM density as 
determined by WMAP without the need for late-decaying moduli fields. 
In fact, the wino component of the $\tz_1$ is usually of order
$0.1-0.2$, so that the $\tz_1$ is still mainly bino-like, but with a 
sufficiently large admixture of wino as to match the WMAP result
on $\Omega_{CDM}h^2$. 
In Sec. \ref{sec:pspace}, we present the parameter space for MWDM,
and show how the assumption of MWDM influences the spectrum of 
sparticle masses.
In Sec. \ref{sec:ddet}, we show rates for direct and indirect 
detection of MWDM. These rates are usually enhanced relative to mSUGRA 
due to the enhanced wino component of the $\tz_1$.
In Sec. \ref{sec:col}, we investigate consequences of MWDM for the 
CERN LHC and the international linear $e^+e^-$ collider (ILC). 
The goal here is to devise a set 
of measurements that can differentiate MWDM from the usual case of 
bino-like DM or MHDM as expected in the mSUGRA model.
For MWDM, 
the neutralino mass gap $m_{\tz_2}-m_{\tz_1}$ is almost always less than
$M_Z$, so that two-body decays of $\tz_2$ are closed, and three 
body decays are dominant. The $m_{\tz_2}-m_{\tz_1}$ mass gap is directly 
measurable at the CERN LHC via the well-known edge in the 
$m(\ell^+\ell^- )$ distribution. The correlation of the $\tz_2 -\tz_1$ 
mass gap against direct and indirect detection rates provides
a distinction between the possible DM candidates. 
Measurements of the $m_{\tz_2}-m_{\tz_1}$ mass gap
at the LHC combined with measurements of chargino and neutralino masses and 
production cross sections as a function of beam polarization
at the ILC would provide the ultimate determination of the presence of 
MWDM in the universe.  
In Sec. \ref{sec:conclude}, we present our conclusions.

\section{Relic density and sparticle mass spectrum}
\label{sec:pspace}

Our goal is to explore SUGRA models with non-universal gaugino masses
leading to MWDM with a neutralino relic density in accord with the WMAP
result. To do so, we adopt the subprogram Isasugra, which is a part of the 
Isajet 7.72 event generator program\cite{isajet}. Isasugra allows
supersymmetric spectra generation using a variety of GUT scale non-universal
soft SUSY breaking terms. The Isasugra spectrum is generated using
2-loop MSSM RGEs for coupling and soft SUSY breaking term evolution.
An iterative approach is used to evaluate the supersymmetric spectrum.
Electroweak symmetry is broken radiatively, so that the magnitude, 
but not the sign, of the superpotential $\mu$ parameter is determined.
The RG-improved 1-loop effective potential is minimized at an optimized scale
which accounts for leading 2-loop terms. Full 1-loop radiative corrections 
are incorporated for all sparticle masses. To evaluate the neutralino 
relic density, we adopt the IsaReD program\cite{isared}, 
which is based on CompHEP
to compute the several thousands of neutralino annihilation and co-annihilation
Feynman diagrams. Relativistic thermal averaging of the cross section
times velocity is performed\cite{gg}. The parameter space we consider is
given by
\be
m_0,\ m_{1/2},\ A_0,\ \tan\beta ,\ sign (\mu ),\ M_1\ {\rm or}\ M_2 ,
\ee
where we take either $M_1$ or $M_2$ to be free parameters, and in general
not equal to $m_{1/2}$.

In Fig. 1, we show our first result. Here, we take $m_0=m_{1/2}=300$ GeV,
with $A_0=0$, $\tan\beta =10$, $\mu >0$ with $m_t=178$ GeV. We plot the 
neutralino relic density $\Omega_{\tz_1}h^2$ in frame {\it a}) versus
variation in the $U(1)$ gaugino mass $M_1$. At $M_1=300$ GeV, 
we are in the mSUGRA case, and $\Omega_{\tz_1}h^2=1.3$, so that the
model would be excluded by WMAP. By decreasing $M_1$, the bino-like
neutralino becomes lighter until two dips in the relic density occur.
These correspond to the cases where $2m_{\tz_1}\simeq m_h$ and $M_Z$
as one moves towards decreasing $M_1$, {\it i.e.} one has either
light Higgs $h$ or $Z$ resonance annihilation. As $M_1$ increases past
its mSUGRA value, the $\tz_1$ becomes increasing wino-like, and the relic 
density decreases. The $W-\tw_{1,2}-\tz_1$ coupling is proportional to the
$SU(2)_L$ gaugino component of the neutralino, 
(and also to the Higgsino components), and so
$\tz_1\tz_1\to W^+W^-$ annihilation becomes enhanced, and the relic 
density is lowered. In this case, the WMAP $\Omega_{\tz_1}h^2$ value is reached
for $M_1=490$ GeV. For still higher $M_1$ values, $\Omega_{\tz_1}h^2$ drops
precipitously, so that other non-neutralino dark matter candidates would 
have to exist to account for the dark matter density in the universe.
In frame {\it b}), we show the bino/wino fraction $R_{\tB ,\tw}$ 
of the $\tz_1$. 
Here, we adopt the notation of Ref. \cite{wss}, wherein the lightest
neutralino is written in terms of its (four component Majorana) 
Higgsino and gaugino components as
\be
\tz_1=v_1^{(1)}\psi_{h_u^0}+v_2^{(1)}\psi_{h_d^0}+v_3^{(1)}\lambda_3 
+v_4^{(1)}\lambda_0 ,
\ee
where $R_{\tw}=|v_3^{(1)}|$ and $R_{\tB}=|v_4^{(1)}|$.
While $R_{\tw}$ increases as $M_1$ increases, its value
when $\Omega_{\tz_1}h^2$ reaches the WMAP value is still only $\sim 0.25$, 
while $R_{\tB}\sim 0.9$. Thus, the $\tz_1$ is still mainly bino-like, with just
enough admixture of wino to give the correct relic density.
This corresponds to the case of MWDM.
A similar plot is obtained by lowering $M_2$, rather than raising $M_1$,
as shown in Fig. \ref{rd_bw2}.
\FIGURE[htb]{
\epsfig{file=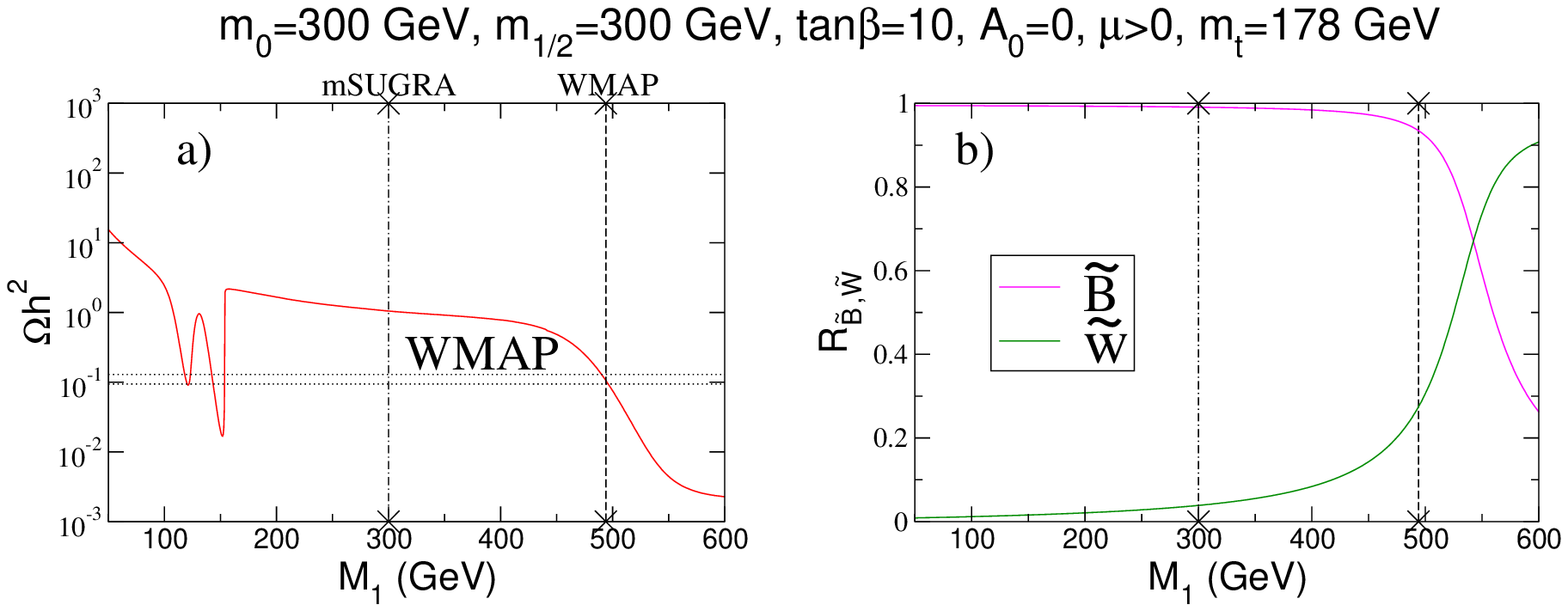,width=6cm,angle=-90} 
\caption{\label{rd_bw}
A plot of {\it a}) relic density $\Omega_{CDM}h^2$ and 
{\it b}) bino/wino component of the lightest neutralino as a
function of $M_1$ for
$m_0=300$ GeV, $m_{1/2}=300$ GeV, $A_0=0$, $\tan\beta =10$, $\mu >0$
and $m_t=178$ GeV.}}
\FIGURE[htb]{
\epsfig{file=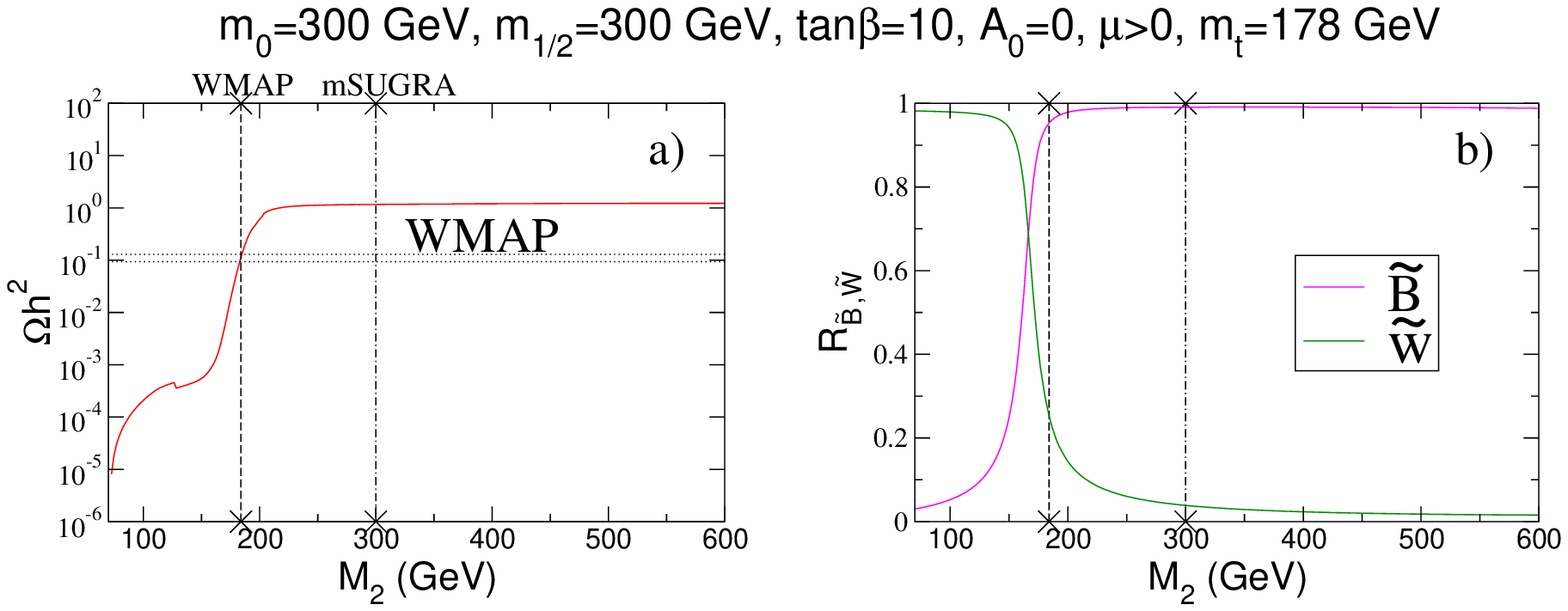,width=6cm,angle=-90} 
\caption{\label{rd_bw2}
A plot of {\it a}) relic density $\Omega_{CDM}h^2$ and 
{\it b}) bino/wino component of the lightest neutralino as a
function of $M_2$ for
$m_0=300$ GeV, $m_{1/2}=300$ GeV, $A_0=0$, $\tan\beta =10$, $\mu >0$
and $m_t=178$ GeV.}}

By raising or lowering the GUT scale gaugino masses in SUGRA models,
the mass of the neutralinos will obviously change since $M_1$ and $M_2$
enter directly into the neutralino mass matrix. However, various other
sparticle masses will also be affected by varying the gaugino masses, 
since these feed into the soft term evolution via the RGEs. 
In Fig. \ref{mass_M1}, we show the variation of the sparticle 
mass spectrum with respect to the GUT scale ratio 
$M_1/m_{1/2}$ for the same parameters
as in Fig. \ref{rd_bw}. When $M_1/m_{1/2}=1$, there is a
relatively large mass gap between $\tz_2$ and $\tz_1$: 
$m_{\tz_2}-m_{\tz_1}=106.7$ GeV. As $M_1$ is increased until
$\Omega_{\tz_1}h^2=0.11$, the mass gap shrinks 
to $m_{\tz_2}-m_{\tz_1} =31.9$ GeV. The light chargino mass $m_{\tw_1}$
remains essentially constant in this case, since $M_2$ remains fixed
at 300 GeV. However, we notice that as $M_1$ increases, the $\te_R$, 
$\tmu_R$ and $\ttau_1$ masses also increase, since $M_1$ feeds into their
mass evolution via RGEs. 
As the coefficient appearing in front of $M_1$ in the RGEs is larger 
(and with the same sign) for the right handed sfermions than 
for the left handed ones, one expects, in general, a departure 
from the usual mSUGRA situation where the lightest sleptons are right-handed.
As a matter of fact, whereas in mSUGRA $m_{\te_L}>> m_{\te_R}$, in the 
case of MWDM, instead, for the particular parameter space slice 
under consideration, we find that $m_{\te_L}\sim m_{\te_R}$.
As shown in the figure, the right-handed squark masses also increase
with increasing $M_1$, although the relative effect is less dramatic
than the case involving sleptons: 
the dominant driving term in the RGEs is, in this case, 
given by $M_3$ (absent in the case of sleptons), 
hence variations in the GUT value of $M_1$ produce milder effects.
\FIGURE[htb]{
\epsfig{file=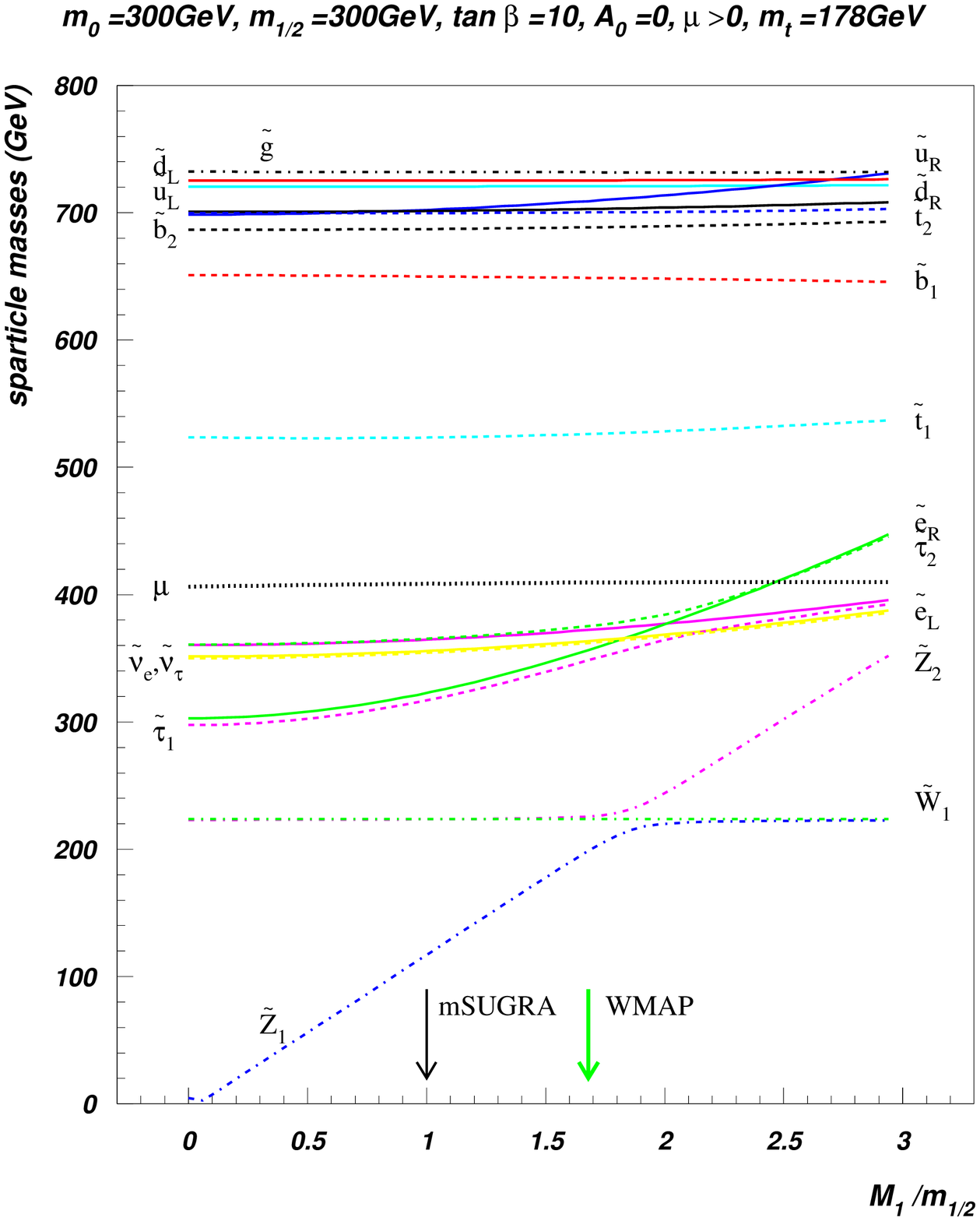,width=12cm} 
\caption{\label{mass_M1}
A plot of various sparticle masses {\it vs.} $M_1/m_{1/2}$ for
$m_0=300$ GeV, $m_{1/2}=300$ GeV, $A_0=0$, $\tan\beta =10$ and $\mu >0$.}}

In Fig. \ref{mass_M2}, we show a plot of sparticle masses for the same
parameters as in Fig. \ref{mass_M1}, but versus $M_2/m_{1/2}$. In this case,
as $M_2$ is decreased from its mSUGRA value of 300 GeV, the
$\tw_1$ and $\tz_2$ masses decrease until $\Omega_{\tz_1}h^2$ reaches 0.11,
where now $m_{\tz_2}-m_{\tz_1}= 22.9$ GeV. In this case, with decreasing
$M_2$, the left- slepton and sneutrino masses also decrease, again
leading to $m_{\te_L}\sim m_{\te_R}$. The left-handed squark masses 
similarly decrease. Right-handed sfermion masses are, instead, not affected, with the net result that the mSUGRA $m_{\te_L}>> m_{\te_R}$ hierarchy is again altered. 
\FIGURE[htb]{
\epsfig{file=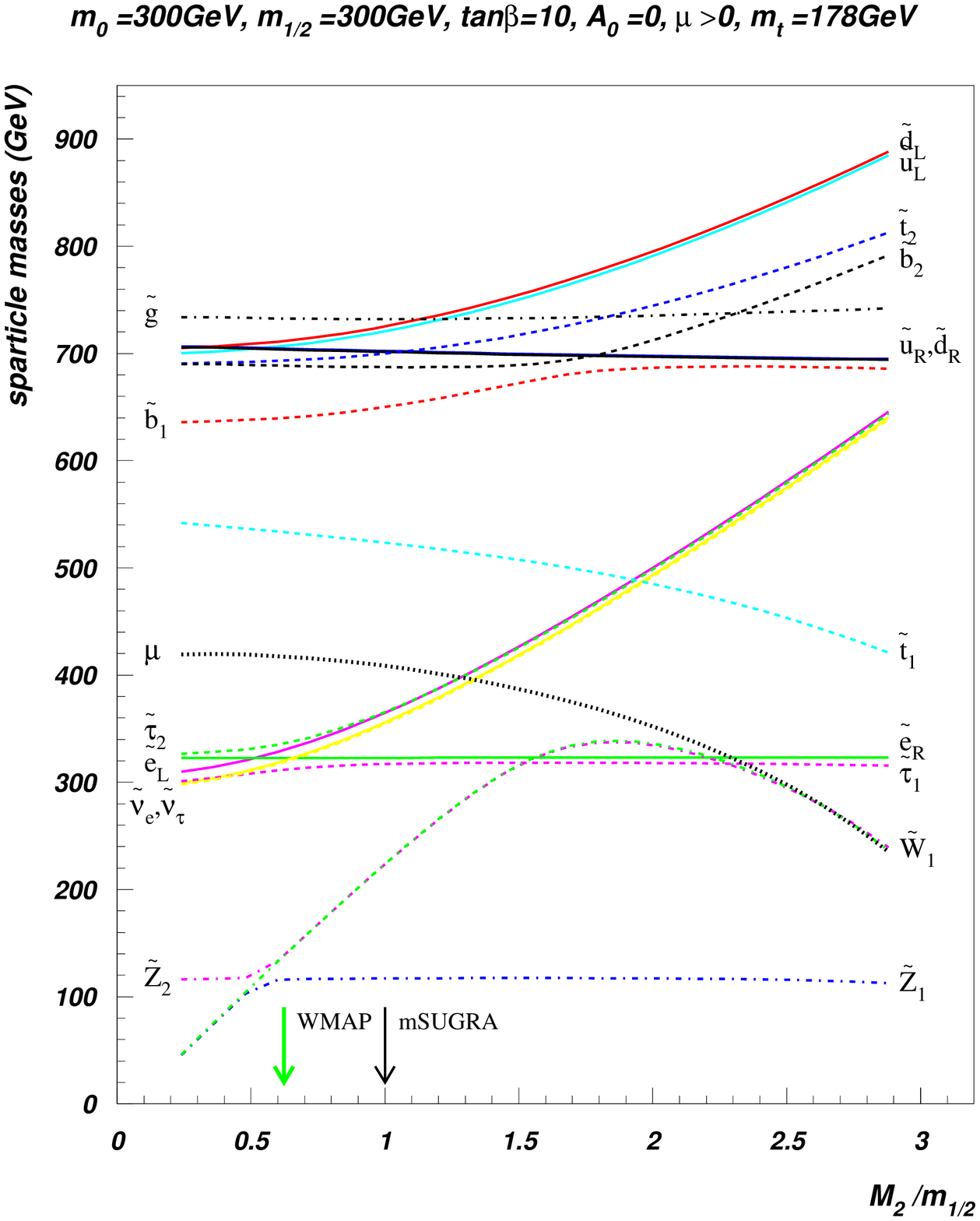,width=12cm} 
\caption{\label{mass_M2}
A plot of various sparticle masses {\it vs.} $M_2/m_{1/2}$ for
$m_0=300$ GeV, $m_{1/2}=300$ GeV, $A_0=0$, $\tan\beta =10$ and $\mu >0$.}}

The effect of varying gaugino masses on the allowed region of 
parameter space is illustrated in Fig. \ref{planes}. Here, in frame
{\it a}), we show the case of the mSUGRA model in the $m_0\ vs.\ m_{1/2}$
plane for $A_0=0$, $\tan\beta =10$ and $\mu >0$. The red shaded regions 
are disallowed by either a stau LSP (left side of plot) or lack of 
REWSB (lower edge of plot). The blue shaded region has a chargino
with mass $m_{\tw_1}<103.5$ GeV, thus violating bounds from LEP2. The dark 
green shaded region has $0.094<\Omega_{\tz_1}h^2<0.129$, in accord with 
the WMAP measurement. The light green shaded region has 
$\Omega_{\tz_1}h^2<0.094$, so that additional sources of dark matter 
would be needed. We see the stau co-annihilation region appearing along 
the left edge of the allowed parameter space, and the bulk region 
appearing at low $m_0$ and low $m_{1/2}$. The $h$ annihilation corridor
appears also at low $m_{1/2}$ along the edge of the LEP2 excluded region.
In frame {\it b}), we take $M_1/m_{1/2}=1.5$, so that the $\tz_1$ becomes more
wino-like. In response, we see that a large new bulk region has appeared
at low $m_0$ and low $m_{1/2}$. In frame {\it c}), we increase 
$M_1/m_{1/2}$ to 1.75. In this case, most of the $m_0\ vs.\ m_{1/2}$ 
plane is now allowed, although much of it has $\Omega_{\tz_1}h^2$
below the WMAP central value for the CDM relic density.
\FIGURE[htb]{
\epsfig{file=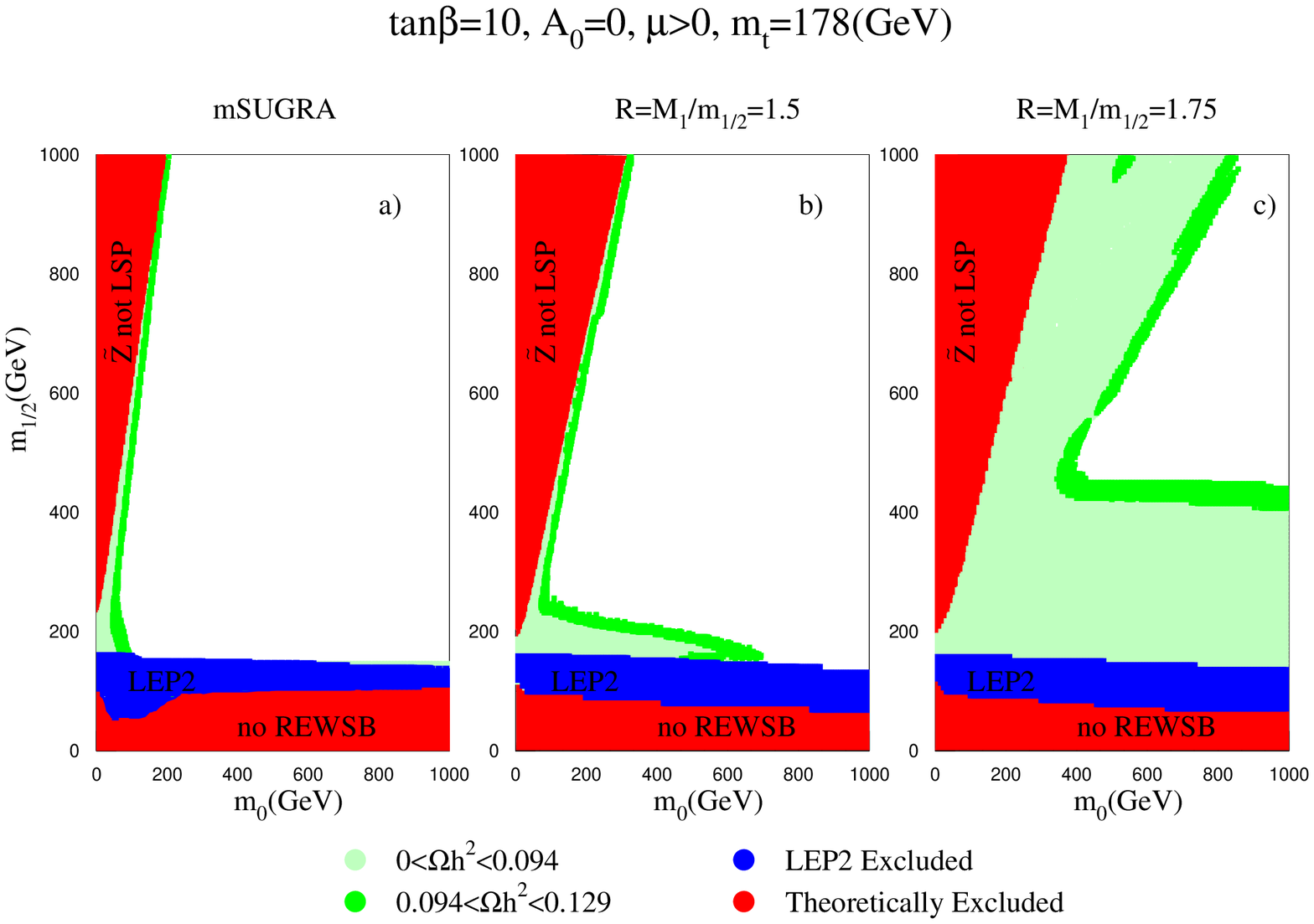,width=13cm} 
\caption{\label{planes}
WMAP allowed regions in the $m_0\ vs.\ m_{1/2}$ plane for
$\tan\beta =10$, $A_0=0$, $\mu >0$ and
{\it a}) $M_1/m_{1/2}=1$, {\it b}) $M_1/m_{1/2}=1.5$ and
{\it c}) $M_1/m_{1/2}=1.75$.}}

It should be apparent now that {\it any} point in the $m_0\ vs.\ m_{1/2}$ plane
can become WMAP allowed by either increasing $M_1$ or decreasing $M_2$ to a
suitable degree as to obtain MWDM. 
To illustrate this, we plot in Fig. \ref{planes_r}
the ratio $r_{1}\equiv M_1/m_{1/2}$ in frame {\it a}) or 
$r_2= M_2/m_{1/2}$ in frame {\it b}) needed to achieve a 
relic density in accord with the WMAP central value. 
We see in frame {\it a}) that $r_1$ increases as one moves from 
lower-left to upper-right, reflecting the greater wino component of $\tz_1$
that is needed to overcome the increasing $\Omega_{\tz_1}h^2$ which is 
expected in the mSUGRA model. We also see on the left side of the plot that
$r_1\le 1$ is allowed, since then $\Omega_{\tz_1}h^2\le 0.11$ 
already in the mSUGRA case. The structure at high $m_{1/2}$ and 
$m_0\sim 400-500$ GeV results because increasing $M_1$ increases
$m_{\tz_1}$ until $2m_{\tz_1}\sim m_A$ and the $A$-funnel begins to come
into effect (even though $\tan\beta $ is small).
\FIGURE[htb]{
\mbox{\hspace{-0.5cm}\epsfig{file=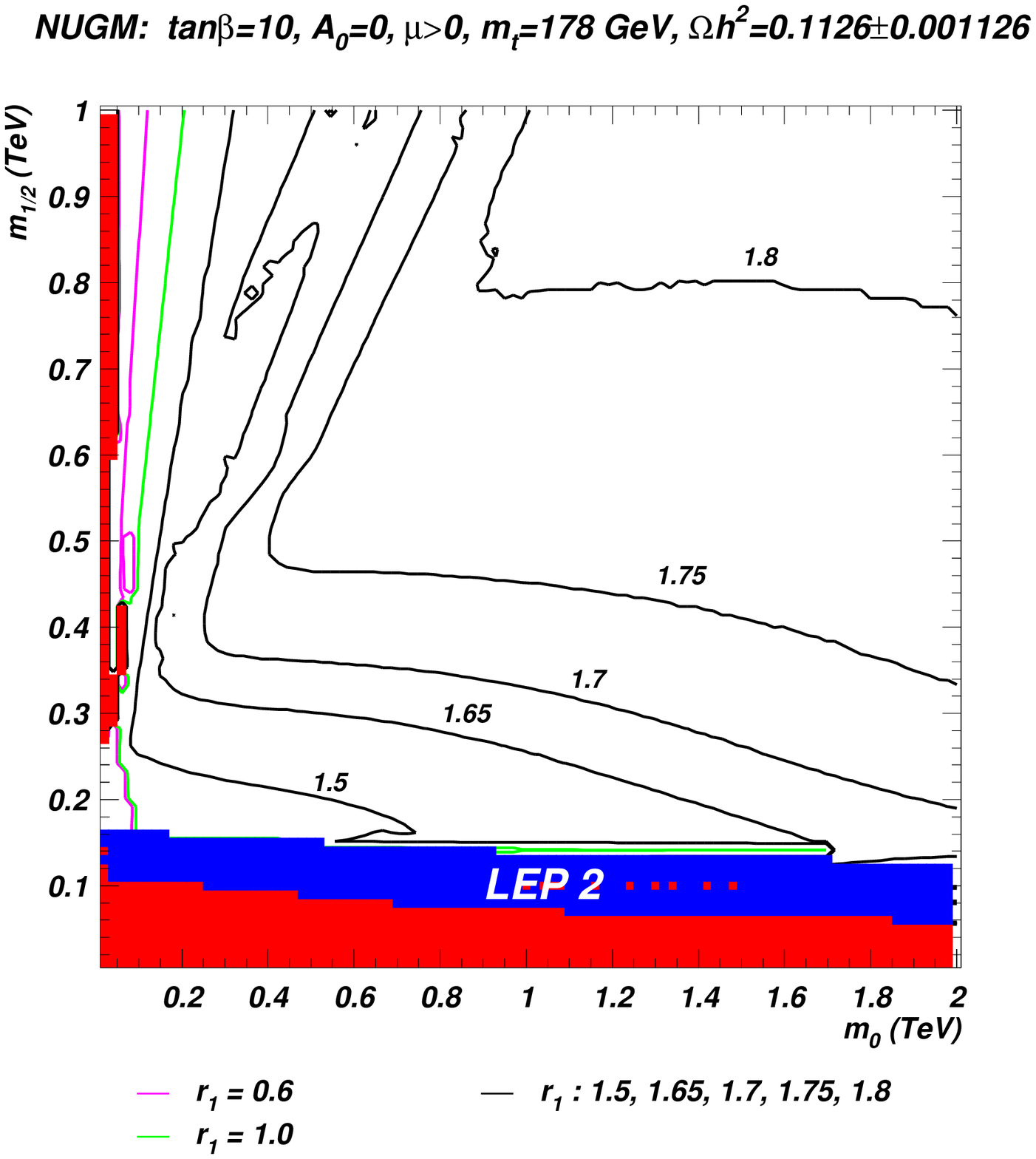,width=8cm} 
\epsfig{file=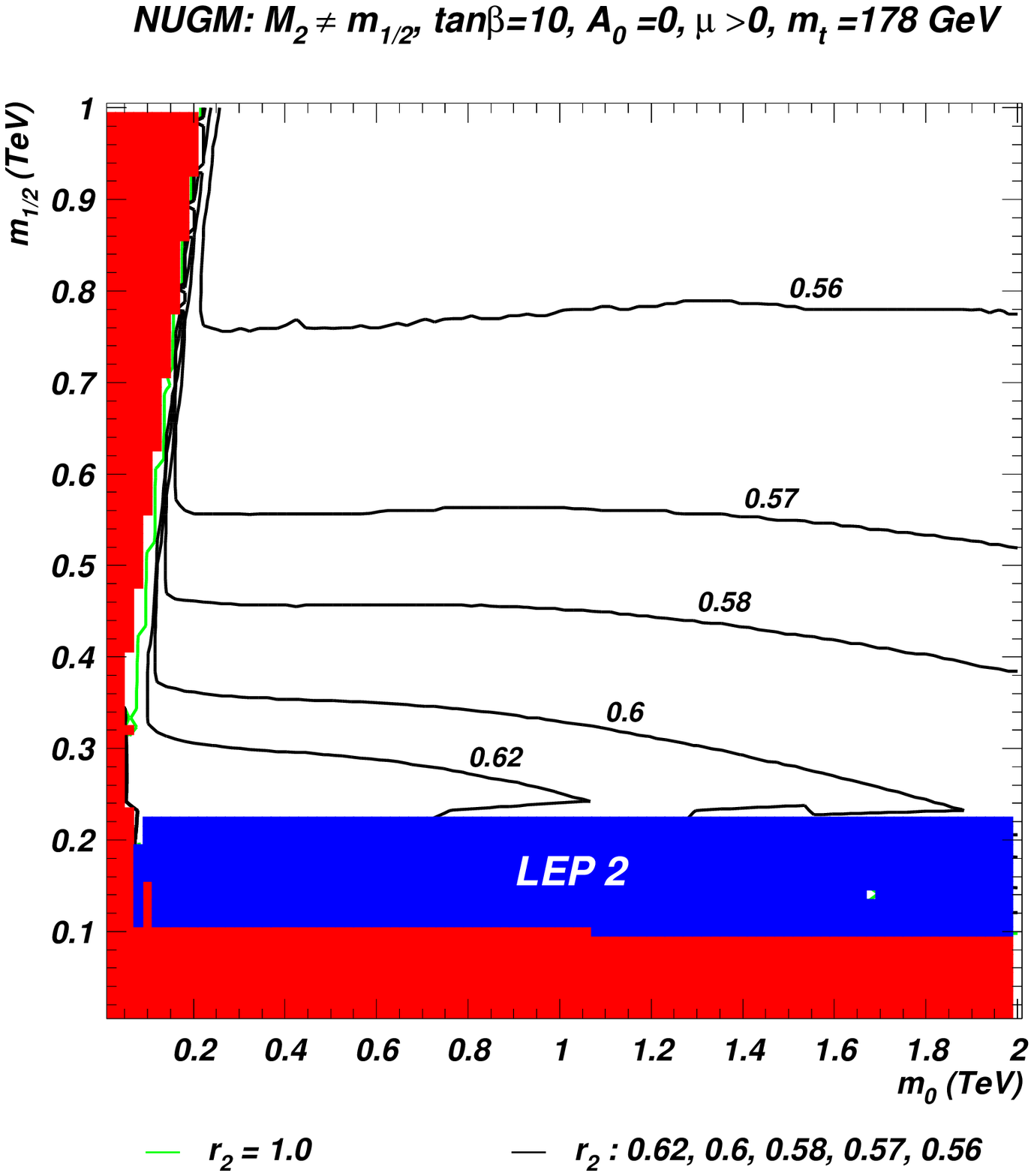,width=8cm}}
\caption{\label{planes_r}
Contours of {\it a}) $r_1$ and {\it b}) $r_2$ in the 
$m_0\ vs.\ m_{1/2}$ plane for
$\tan\beta =10$, $A_0=0$, $\mu >0$. Each point has
$\Omega_{\tz_1}h^2 =0.11$.}}

\section{Direct and indirect detection of mixed wino dark matter}
\label{sec:ddet}

In this section, we turn to consequences of MWDM for direct and indirect 
detection of neutralino dark matter\cite{eigen,fmw}. 
We adopt the DarkSUSY code\cite{darksusy}, 
interfaced to Isajet, for the computation of the various rates, and resort 
to the Adiabatically Contracted 
N03 Halo model\cite{n03} for the dark matter distribution 
in the Milky Way\footnote{For a comparison of the implications of 
different halo model choices for indirect DM detection rates, see
{\it e.g.} Refs. \cite{bo,bbko,antimatter,nuhm}.}.
We evaluate the following neutralino DM detection rates:
\begin{itemize}
\item Direct neutralino detection via 
underground cryogenic detectors\cite{direct}.
Here, we compute the spin independent neutralino-proton scattering 
cross section, and compare it to expected sensitivities\cite{bbbo} for Stage 2 
detectors (CDMS2\cite{cdms2}, Edelweiss2\cite{edelweiss}, 
CRESST2\cite{cresst}, ZEPLIN2\cite{zeplin}) 
and for Stage 3, ton-size detectors (XENON\cite{xenon}, 
Genius\cite{genius}, ZEPLIN4\cite{zeplin4} and WARP\cite{warp}). 
We take here as benchmark experimental reaches of Stage 2 and Stage 3 
detectors the projected sensitivities of, respectively, 
CDMS2 and XENON 1-ton at the corresponding neutralino mass.

\item Indirect detection of neutralinos via neutralino annihilation to
neutrinos in the core of the Sun\cite{neut_tel}. 
Here, we present rates for detection of $\nu_\mu \to \mu$ conversions
at Antares\cite{antares} or IceCube\cite{icecube}. 
The reference experimental sensitivity we use is that of IceCube, 
with a muon energy threshold of 25 GeV, corresponding to a flux 
of about 40 muons per ${\rm km}^2$ per year. 
\item Indirect detection of neutralinos via neutralino annihilations in the
galactic center leading to gamma rays\cite{gammas}, 
as searched for by EGRET\cite{egret}, and 
in the future by GLAST\cite{glast}. 
We evaluate the integrated continuum $\gamma$ ray flux above a 
$E_\gamma=1$ GeV threshold, and assume a GLAST sensitivity 
of 1.0$\times10^{-10}\ {\rm cm}^{-2}{\rm s}^{-1}$.
\item Indirect detection of neutralinos via neutralino annihilations in the
galactic halo leading to cosmic antiparticles, including
positrons\cite{positron} (HEAT\cite{heat}, Pamela\cite{pamela} 
and AMS-02\cite{ams}), antiprotons\cite{pbar} (BESS\cite{bess}, 
Pamela, AMS-02) and anti-deuterons ($\bar{D}$s) (BESS\cite{bessdbar}, 
AMS-02, GAPS\cite{gaps}). 
For positrons and antiprotons we evaluate the averaged differential 
antiparticles flux in a projected energy bin centered at a kinetic 
energy of 20 GeV, where we expect an optimal statistics and 
signal-to-background ratio at space-borne antiparticles 
detectors\cite{antimatter,statistical}. We use as benchmark 
experimental sensitivity that of the Pamela experiment 
after three years of data-taking. 
Finally, the average differential antideuteron flux has been 
computed in the $0.1<E_{\bar D}<0.4$ GeV range, 
and compared to the estimated GAPS sensitivity\cite{gaps}.
\end{itemize}

In Fig. \ref{dmrates1}, we show various direct and indirect DM detection 
rates for $m_0=m_{1/2}=300$ Gev, with $A_0=0$, $\tan\beta =10$ and $\mu >0$,
while $M_1$ is allowed to vary. The $M_1$ value corresponding to the mSUGRA model is 
denoted by a dot-dashed vertical line, while the one where 
$\Omega_{\tz_1}h^2=0.11$ by a dashed vertical line denoted 
WMAP. 

In frame {\it a}), we plot the spin-independent neutralino-proton
scattering cross section. Both the squark-mediated 
and Higgs mediated neutralino-proton scattering amplitudes are 
enhanced by more than one order of magnitude due to the 
increasing wino nature of the $\tz_1$. 
The reason for the enhancement is traced back to the structure 
of the neutralino-quark-squark and neutralino-neutralino-Higgs couplings, 
where the wino fraction is weighed by the $SU(2)$ coupling, 
while the bino fraction by the (smaller) $U(1)$ coupling.
\FIGURE[htb]{
\mbox{\hspace{-1cm}\epsfig{file=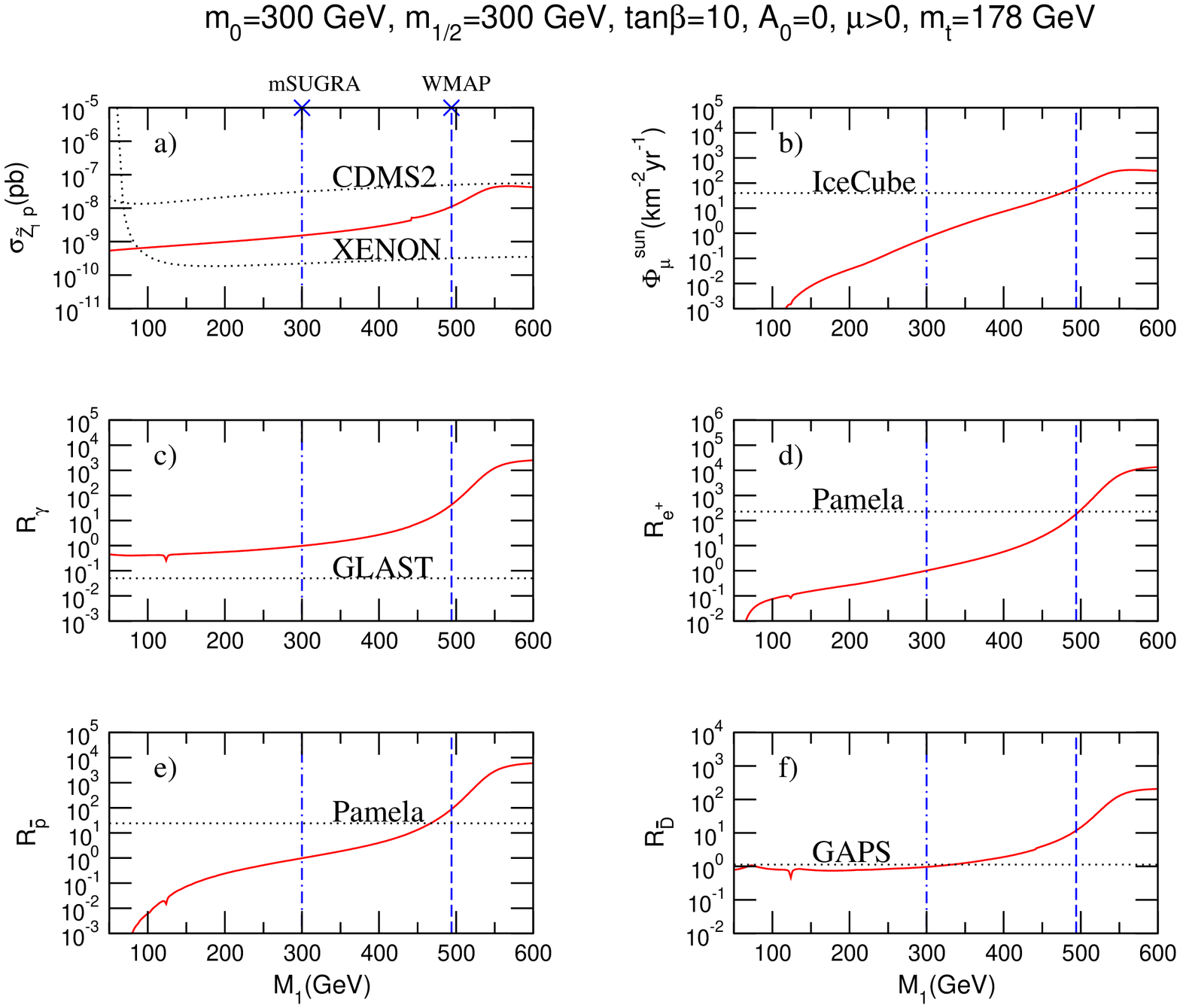,width=14cm,angle=-90} }
\caption{\label{dmrates1}
Rates for direct and indirect detection of neutralino dark matter
vs. $M_1$ for $m_0=m_{1/2}=300$ GeV, with
$\tan\beta =10$, $A_0=0$, $\mu >0$. 
Frames {\it c}) -{\it f}) show the ratio of indirect detection rates 
compared to the mSUGRA model. 
In this plot, we adopt the 
N03 distribution for halo dark matter.}}

In frame {\it b}), we show the flux of muons from neutralino 
pair annihilations  in the core of the Sun.
While the muon flux is below the reach of IceCube in the mSUGRA case, it has
climbed into the observable region when the $\tz_1$ has become sufficiently
wino-like as to fulfill the WMAP measured DM relic density.

In frames {\it c}), {\it d}), {\it e}) and {\it f}) we show the
flux of photons, positrons, 
antiprotons and antideuterons, respectively. 
The results here are plotted as ratios of fluxes normalized to the mSUGRA point, 
in order to give results that are approximately halo-model independent. 
(We do show the above described expected experimental reach lines as obtained by using
the Adiabatically Contracted 
N03 Halo model\cite{n03}.) All
rates are enhanced, with respect to the mSUGRA case, 
by 2 to 3 orders of magnitude, due to the increasing cross section for
$\tz_1\tz_1\to W^+W^- $ annihilation in the galactic halo. 
In particular, antimatter fluxes are always below future sensitivities 
for the mSUGRA setup, while they all rise to a detectable level when the WMAP point is reached.   

In Fig. \ref{dmrates2}, we show the same direct and indirect DM detection
rates as in Fig. \ref{dmrates1}, except this time versus $M_2$ instead of
$M_1$. In this case, the various rates are all increasing as $M_2$ decreases,
entering the region of MWDM. Indirect detection rates again feature 
enhancements as large as 2 orders of magnitude with respect to the mSUGRA 
scenario, when the WMAP relic abundance is reached. The abrupt decrease 
in the rates below $M_2\sim100$ GeV is due, instead, to the $m_{\tz_1}<m_W$ threshold.
\FIGURE[htb]{
\mbox{\hspace{-1cm}\epsfig{file=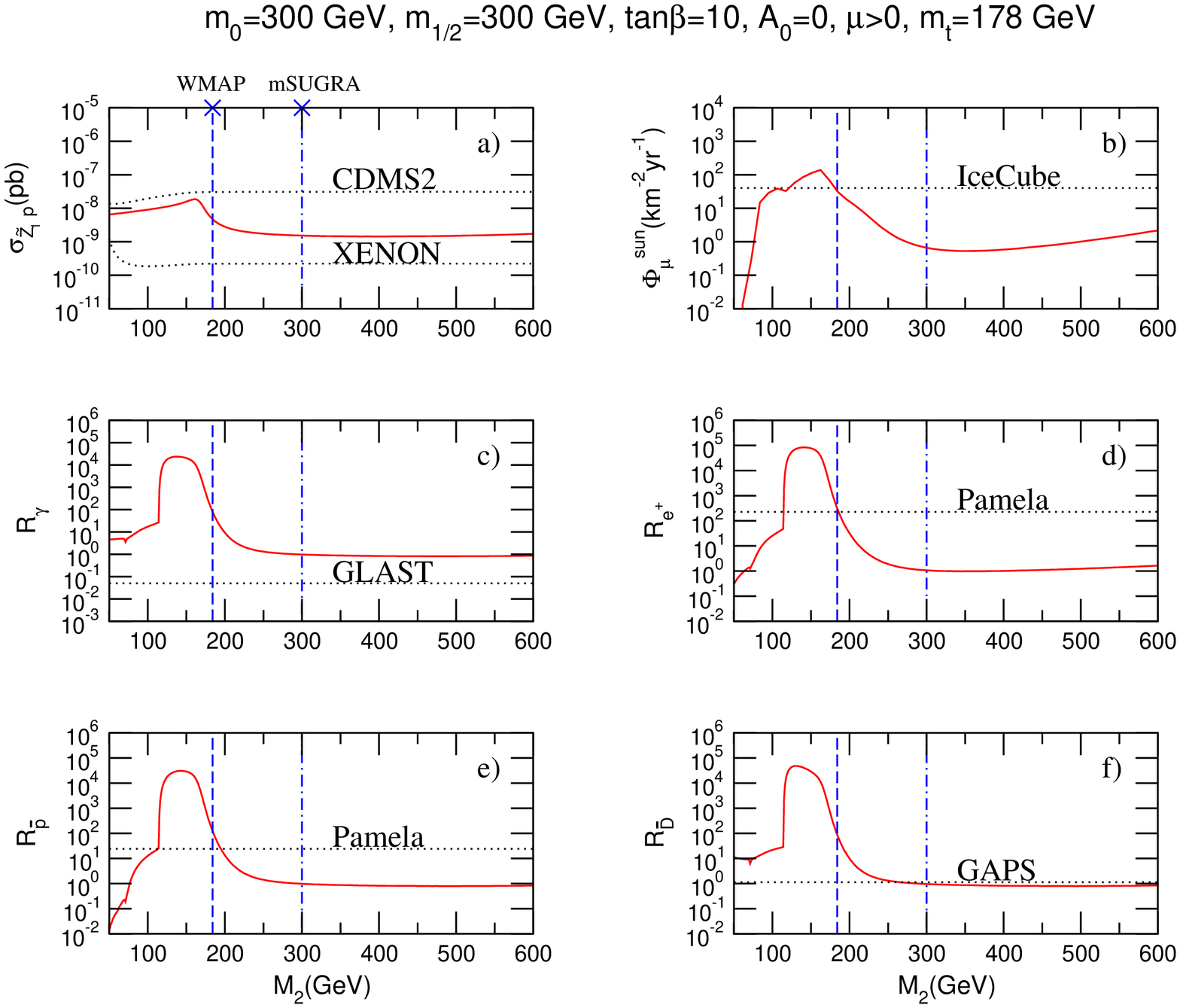,width=14cm,angle=-90} }
\caption{\label{dmrates2}
Rates for direct and indirect detection of neutralino dark matter
vs. $M_2$ for $m_0=m_{1/2}=300$ GeV, with
$\tan\beta =10$, $A_0=0$, $\mu >0$. 
Frames {\it c}) -{\it f}) show the ratio of indirect detection rates 
compared to the mSUGRA model. 
In this plot, we adopt the 
N03 distribution for halo dark matter.}}

In Fig. \ref{dmplanes}, we show regions of the $m_0\ vs.\ m_{1/2}$ plane
for $A_0=0$, $\tan\beta =10$ and $\mu >0$ which are accessible to 
various direct and indirect DM search experiments. 
The visibility criteria we adopt here follow the same approach 
outlined in Ref.~\cite{nuhm}. 
The gray shaded regions in the plots are already excluded, 
at 95\% C.L., by a $\chi^2$ analysis of the computed signal plus background 
$\bar{p}$ flux compared to the available antiprotons data 
(for details see\cite{antimatter}).
Observable rates for $\gamma$ detection by GLAST occur throughout all three planes,
due to the high DM density assumed at the galactic core in the N03 halo model.
In frame {\it a}), we show the case of the mSUGRA model. Only 
small regions at low $m_0$ and low $m_{1/2}$ are accessible to $\bar{D}$
searches by GAPS and $\bar{p}$ searches by Pamela. A tiny region is also 
accessible to CDMS2, and a much larger region is accessible to Stage 3 direct 
detection experiments such as XENON. In frame {\it b}), we increase 
$M_1(M_{GUT})$ at every point in the plane as in Fig. \ref{planes_r} 
until $\Omega_{\tz_1}h^2=0.11$. The corresponding neutralino masses are 
therefore accordingly increased with respect to the mSUGRA case. 
Nevertheless, we see that the regions accessible to direct and indirect DM detection 
have vastly increased. The $\bar{D}$ search by GAPS can cover 
$m_{1/2}\lesssim 400-500$ GeV. The $e^+$ and $\bar{p}$ searches by Pamela
can see to $m_{1/2}\sim 250$ GeV and 350 GeV, respectively.
In addition, a region has opened up which is accessible to IceCube searches
for dark matter annihilation in the core of the Sun. The Stage 3 dark matter
detectors can see most of the $m_0\ vs.\ m_{1/2}$ plane, with the 
exception of the region at large $m_{1/2}$ and low $m_0$ where a much lower
wino component of the $\tz_1$ is required to bring the relic density
into line with the WMAP measurement (here, early universe $\tz_1\tz_1$ 
annihilations are already somewhat enhanced by the proximity of the 
$A$-pole and the stau co-annihilation region).
In frame {\it c}), we show again the $m_0\ vs.\ m_{1/2}$ plane, 
but this time we have {\it reduced} $M_2$ until the $\Omega_{\tz_1}h^2=0.11$
value is reached. Again, many of the direct and indirect detection regions 
are expanded compared to the mSUGRA case. 
We remark that, although in this last case the neutralino mass is lower 
than in the case shown in frame {\it b}), direct detection and neutrino fluxes 
are somewhat less favored. This depends on the relative higgsino fraction, 
which critically enters in the neutralino-proton scattering cross section 
as well as in the neutralino capture rate in the Sun: 
raising $M_1$ shifts the gaugino masses closer to $\mu$, 
hence increasing the higgsino fraction and the resulting neutralino cross sections off matter.
\FIGURE[htb]{
\mbox{\epsfig{file=msugra_10_def.eps,width=8.5cm}\vspace*{1cm}}
\mbox{\hspace*{-1cm}\epsfig{file=mwdm_r1_10.eps,width=8.5cm} 
\epsfig{file=mwdm_r2_10.eps,width=8.5cm}}
\caption{\label{dmplanes}
Regions of visibility for direct and indirect dark matter searches
in the $m_0\ vs.\ m_{1/2}$ plane for
$\tan\beta =10$, $A_0=0$, $\mu >0$. The upper frame {\it a}) shows the mSUGRA model, while frame {\it b}) corresponds to the MWDM model with non-universal $M_1$ and frame {\it c}) with non-universal $M_2$. In this plot, we adopt the 
Adiabatically Contracted N03 Halo Model for the galactic dark matter
distribution. For this halo model, detection of $\gamma$s by GLAST should occur
over all three planes.}}

\section{Mixed wino dark matter at colliders}
\label{sec:col}

An important question is whether collider experiments would be able to 
distinguish the case of MWDM from other forms of neutralino DM
such as bino-DM or MHDM as occur in the mSUGRA model. We have seen 
from the plots of sparticle mass spectra that the squark and
gluino masses vary only slightly with changing $M_1$ or $M_2$. However, the
chargino and neutralino masses change quite a bit, and in fact
rather small mass gaps $m_{\tw_1}-m_{\tz_1}$ and $m_{\tz_2}-m_{\tz_1}$ are 
in general expected in the case of MWDM as compared with the case from models
containing gaugino mass unification.

In Fig. \ref{z21gap}, we show contours of 
the mass gap $m_{\tz_2}-m_{\tz_1}$ in the $m_0\ vs.\ m_{1/2}$ plane for 
$A_0=0$, $\tan\beta =10$ and $\mu >0$ for
{\it a}) the mSUGRA model, {\it b}) the case of MWDM where $M_1$ is
raised at every point until $\Omega_{\tz_1}h^2\to 0.11$ and {\it c})
the case of MWDM where $M_2$ is lowered until $\Omega_{\tz_1}h^2\to 0.11$.
In the case of the mSUGRA model, most of the parameter space has
$m_{\tz_2}-m_{\tz_1}>90$ GeV, which means that $\tz_2\to \tz_1 Z^0$ decay is
allowed. When this decay is allowed, its branching fraction is always large, 
unless it competes with other two-body decays such as $\tz_2\to \tz_1 h$
or $\tz_2\to \bar{f}\tilde f$ or $f{\bar{\tilde f}}$ (where $f$ is 
a SM fermion). In the case of MWDM in frames {\it b}) and {\it c}), 
we see that (aside from the left-most portion of frame {\it b}), which is not a region of MWDM), 
the mass gap is much smaller, so that 
two-body decays of $\tz_2$ and $\tw_1$ are closed and three-body decays
are dominant.
\FIGURE[htb]{
\epsfig{file=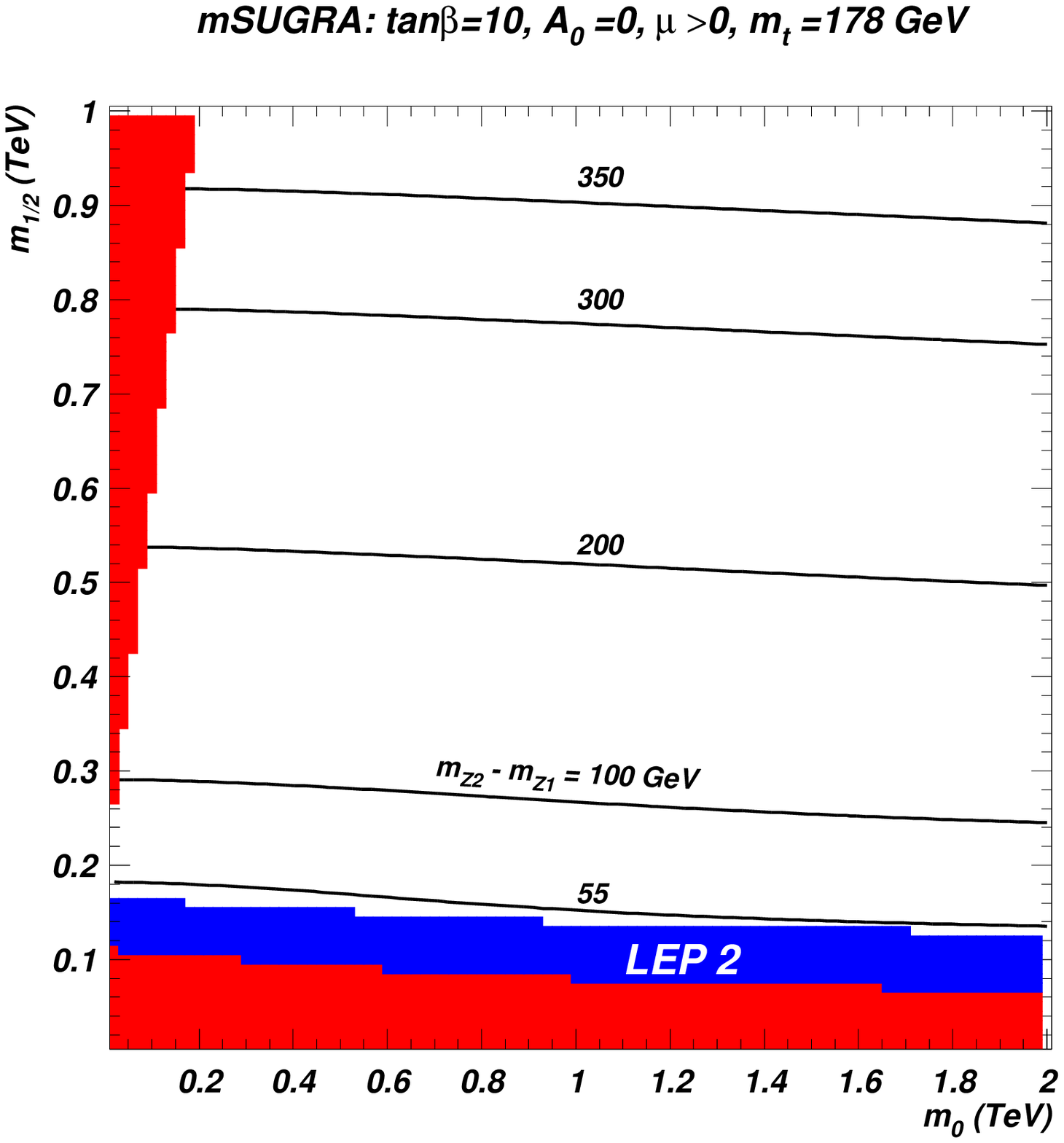,width=8.5cm} 
\mbox{\hspace*{-1cm}\epsfig{file=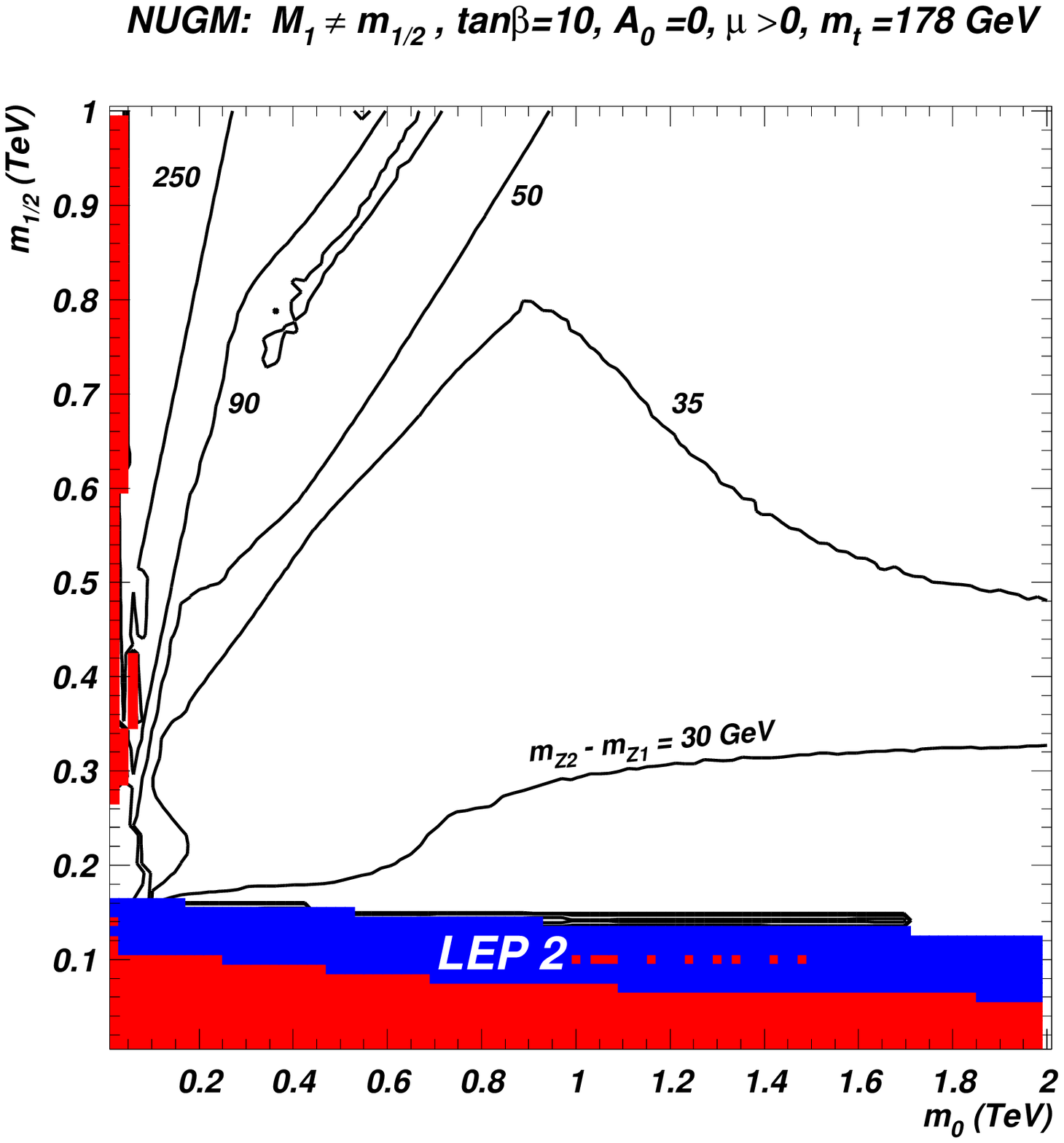,width=8.5cm} 
\epsfig{file=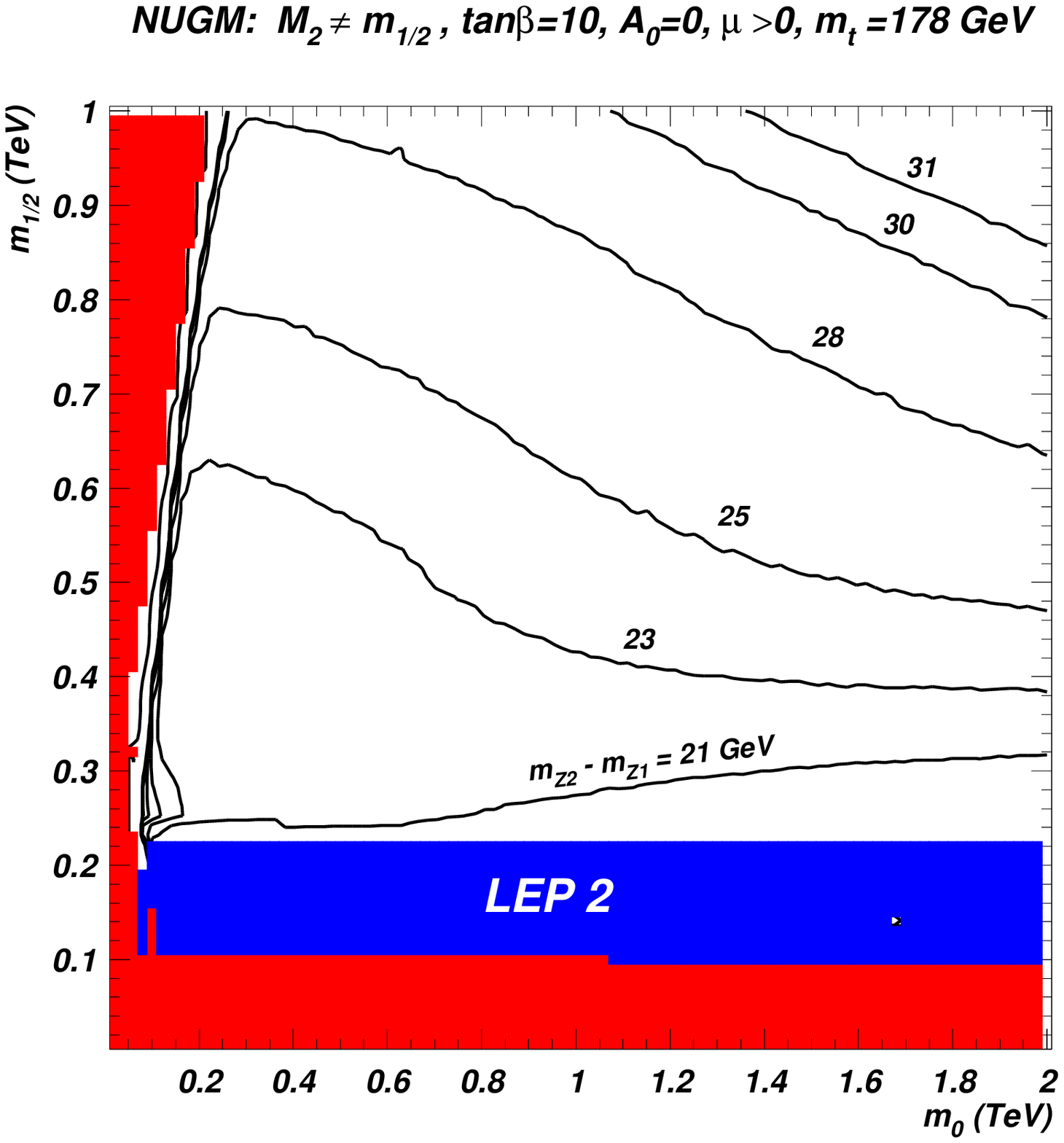,width=8.5cm} }
\caption{\label{z21gap}
Contours of $m_{\tz_2}-m_{\tz_1}$ mass gap in 
the $m_0\ vs.\ m_{1/2}$ plane for
$\tan\beta =10$, $A_0=0$, $\mu >0$ and
{\it a}) mSUGRA model, {\it b}) $M_1>m_{1/2}$ MWDM and
{\it c}) $M_2<m_{1/2}$ MWDM.}}

When the decays 
$\tz_2\to\tell\bar{\ell},\ \bar{\tell}\ell\to \tz_1\ell\bar{\ell}$ 
or $\tz_2\to \tz_1 \ell\bar{\ell}$
are open ($\ell =e$ or $\mu$), 
then prospects are good for measuring the $\tz_2 -\tz_1$ mass gap 
at the CERN LHC and possibly at the Fermilab Tevatron. 
If $\tz_2$'s are produced at large rates either directly or via gluino or
squark cascade decays\cite{cascade}, it should be possible to
identify opposite sign/ same flavor dilepton pairs, to reconstruct their
invariant mass, and extract the upper edge of the 
invariant mass distribution\cite{mlldist}.
In Fig. \ref{bfs}, we show the branching fraction 
$BF(\tz_2\to \tz_1 e^+e^-)$ versus $M_1$ (left-side) or versus
$M_2$ (right-side) for a variety of
choices of $m_0$, $m_{1/2}$ and $\tan\beta$. The mSUGRA model value
is denoted by the dot-dashed vertical line, while the $M_{1,2}$ value
at which $\Omega_{\tz_1}h^2\to 0.11$ is indicated by the dotted
vertical line. As one moves to higher $M_1$ (or lower $M_2$) values, in most cases
the leptonic three-body decays of $\tz_2$ become enhanced, usually 
because as $M_1$ grows ($M_2$ decreases), the two-body decay modes become kinematically closed,
and only three-body decays are allowed.
Thus, while the mSUGRA model yields large rates for $\tz_2\to \tz_1 e^+e^-$
only when $m_{1/2}\alt 220$ GeV, this decay mode is almost always open in the
case of MWDM. The only exception occurs when the stau co-annihilation or the
$A$-funnel act to lower the relic density, so that a large $M_1$ or small
$M_2$ is not needed to obtain the correct relic density; this, however, is not the case
of MWDM. 
\FIGURE[htb]{
\mbox{\hspace*{-1.5cm}\epsfig{file=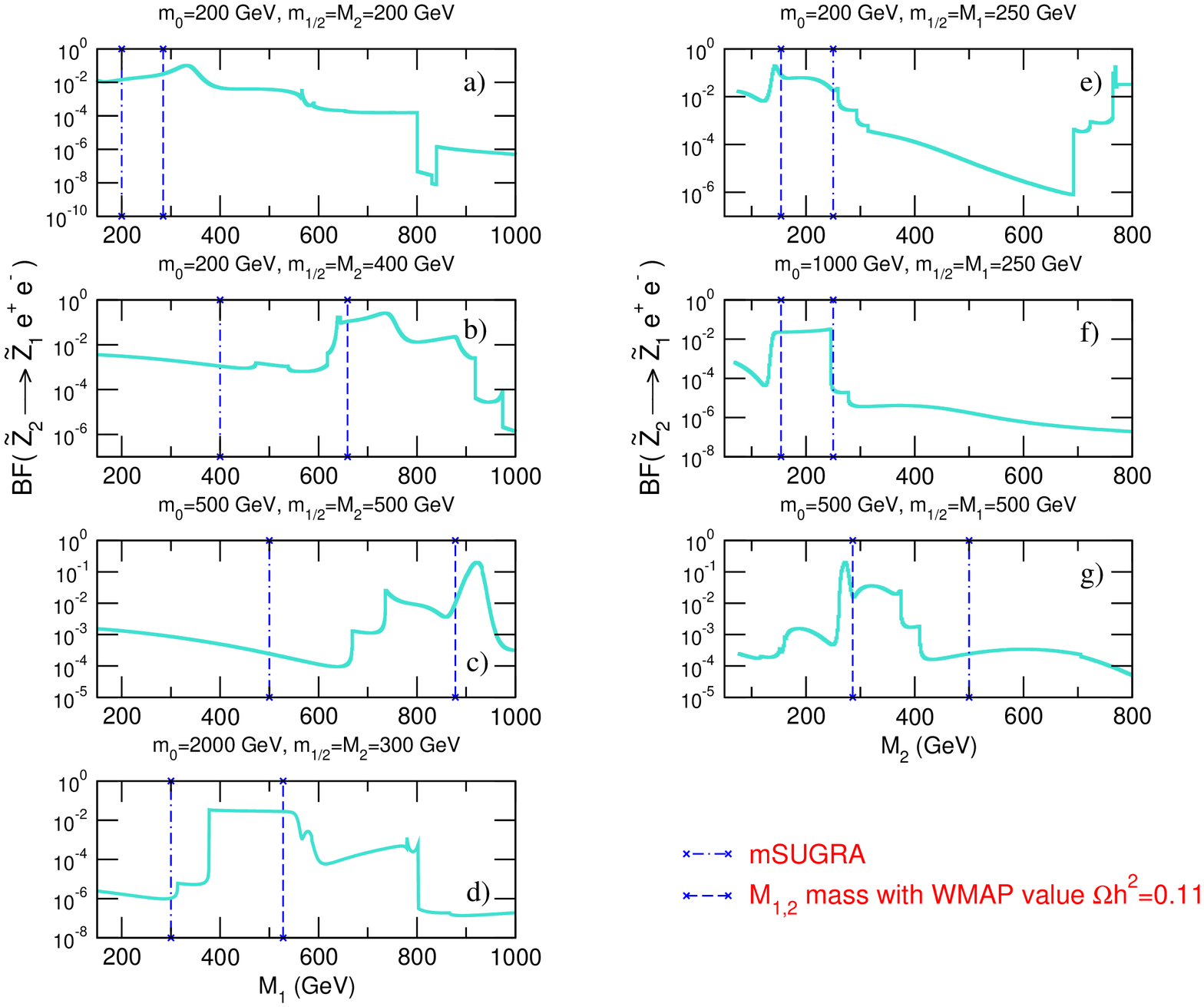,width=14cm,angle=-90}}
\caption{\label{bfs}
The branching fraction for $\tz_2\to\tz_1 e^+e^-$ decay is plotted vs.
$M_1$ (left-side) or $M_2$ (right-side) for various points in the 
MWDM model parameter space.
The $M_{1,2}$ value from mSUGRA is denoted by the dot-dashed lines, while
the $M_{1,2}$ value which gives $\Omega_{\tz_1}h^2=0.11$ is indicated by
dotted lines.
}}

\subsection{CERN LHC}
\label{ssec:lhc}

If the $R$-parity conserving MSSM is a good description of nature
at the weak scale, then
multi-jet plus multi-lepton plus $\eslt$ events should occur at
large rates at the CERN LHC, provided that 
$m_{\tg}\alt 2-3$ TeV.
The LHC reach for SUSY in the mSUGRA model has been calculated in
Ref. \cite{susylhc}. The mSUGRA reach results should also 
apply qualitatively to the MWDM case, since the values of 
$m_{\tg}$ and $m_{\tq}$ change little in going from mSUGRA to MWDM,
and the reach plots mainly depend on these masses.

For SUSY searches at the CERN LHC, Hinchliffe {\it et al.} have 
pointed out\cite{frank} that an approximate value of $m_{\tq}$ or
$m_{\tg}$ can be gained by extracting the maximum in the
$M_{eff}$ distribution, where 
$M_{eff}=\eslt +E_T(jet\ 1)+E_T(jet\ 2)+E_T(jet\ 3)+E_T(jet\ 4)$.
This statement holds true in models with MWDM, as well as in 
models with gaugino mass unification, so that the approximate
mass scale of strongly interacting sparticles will be known soon 
after a supersymmetry signal has been established.  

In mSUGRA, a dilepton mass edge should be
visible in SUSY signal events only if $m_{1/2}\alt 250$ GeV 
or if $\tz_2\to \tell\bar{\ell},\ \bar{\tell}\ell$ decays are allowed.
In the case of MWDM, the dilepton mass edge should be visible over almost all 
parameter space. We illustrate the situation for four case studies
listed in Table \ref{tab:mwdm}. The first case, labeled mSUGRA,
has $m_0=m_{1/2}=300$ GeV, with $A_0=0$, $\tan\beta =10$ and $\mu >0$.
In this case, $\tg\tg$, $\tg\tq$ and $\tq\tq$ production occurs with a 
combined cross section of about 12 pb, 
while the total SUSY cross section is around
13.4 pb (the additional 1.4 pb comes mainly from -ino pair production and
-ino-squark or -ino-gluino associated production). 
The case of MWDM1, with $M_1=490$ GeV, has similar rates of 
sparticle pair production. The case of MWDM2, 
with lighter chargino and neutralino masses, has a total production cross 
section of 19.2 pb, wherein strongly interacting sparticles are 
pair produced at similar rates as in mSUGRA or MWDM1, but -ino pairs are 
produced at a much larger rate $\sim 6.1$ pb.
We also show a case of MHDM from the HB/FP region of the mSUGRA model as
an alternative low $\tz_2 -\tz_1$ mass gap model to compare against MWDM
scenarios.
%
\begin{table}
\begin{tabular}{lcccc}
\hline
parameter & mSUGRA & MWDM1 & MWDM2 & MHDM \\
\hline
$M_1$ & 300 & 490 & 300 & 300 \\
$M_2$ & 300 & 300 & 187 & 300 \\
$\mu$ & 409.2 & 410.1 & 417.8 & 166.1 \\
$m_{\tg}$ & 732.9 &  732.8 & 733.0 & 854.6 \\
$m_{\tu_L}$ & 720.9 & 721.1 & 706.9 & 3467.2 \\
$m_{\tst_1}$ & 523.4 & 526.0 & 533.2 & 2075.8 \\
$m_{\tb_1}$ & 650.0 & 648.9 & 640.2 & 2847.0 \\
$m_{\te_L}$ & 364.7 & 371.7 & 330.0 & 3449.7 \\
$m_{\te_R}$ & 322.8 & 353.7 & 322.7 & 3449.4 \\
$m_{\tw_2}$ & 432.9 & 433.8 & 435.9 & 288.9 \\
$m_{\tw_1}$ & 223.9 & 224.0 & 138.3 & 146.6 \\
$m_{\tz_4}$ & 433.7 & 435.7 & 436.2 & 296.9 \\
$m_{\tz_3}$ & 414.8 & 415.6 & 424.1 & 179.0 \\ 
$m_{\tz_2}$ & 223.7 & 225.4 & 138.8 & 159.2 \\ 
$m_{\tz_1}$ & 117.0 & 193.5 & 115.9 & 101.5 \\ 
$m_A$ & 538.6 & 544.1 & 523.6 & 3409.9 \\
$m_{H^+}$ & 548.0 & 553.5 & 533.1 & 3433.3 \\
$m_h$ & 115.7 & 115.8 & 115.3 & 118.9 \\
$\Omega_{\tz_1}h^2$& 1.3 & 0.11 & 0.11 & 0.13 \\
$BF(b\to s\gamma)$ & $3.2\times 10^{-4}$ & $3.2\times 10^{-4}$ &
$3.3\times 10^{-4}$ & $3.4\times 10^{-4}$ \\
$\Delta a_\mu    $ & $12.1 \times  10^{-10}$ & $11.8 \times  10^{-10}$ & 
$15.9\times 10^{-10}$ & $3.9\times 10^{-11}$ \\ 
$\sigma_{sc} (\tz_1p )$ & 
$2.6\times 10^{-8}\ {\rm pb}$ & $2.2\times 10^{-7}\ {\rm pb}$ & 
$7.1\times 10^{-8}\ {\rm pb}$ & $1.8\times 10^{-8}\ {\rm pb}$\\
\hline
\end{tabular}
\caption{Masses and parameters in~GeV units
for mSUGRA, MWDM and MHDM models. In the first three cases,
$m_0=m_{1/2} =300$ GeV, $A_0=0$, $\tan\beta =10$ and $m_t=178$ GeV.
The case of MHDM has the same parameters, except $m_0=3451.8$ GeV,
with $m_t=175$ GeV.
}
\label{tab:mwdm}
\end{table}

We have generated 50K 
LHC SUSY events for each of these cases using Isajet 7.72, 
and passed them through 
a toy detector simulation. The toy detector is divided into 
calorimeter cells of
size $\Delta\eta \times\Delta\phi =0.05\times 0.05$ extending out to
$|\eta |<5$, with no transverse shower spreading. We invoke EM smearing with
$3\%/\sqrt{E}+.5\%$, hadronic smearing with $80\%/\sqrt{E}+3\%$ out
to $|\eta |=2.6$,
and forward calorimeter hadronic smearing with $100\%/\sqrt{E} +5\%$.
Jets are clustered using a UA1 type algorithm with cone size 
$R=\sqrt{\Delta\eta^2+\Delta\phi^2}=0.7$, with $E_{jet}(min)=25$ GeV.
Leptons ($\ell =e\ {\rm or}\ \mu$) with $E_\ell >10$ GeV are classified 
as isolated if $E_T(cone)<5$ GeV in a cone of $R=0.3$ about the lepton's 
direction. Since gluino and squark  masses of the three case studies 
are similar to those 
of LHC point 5 of the study of Hinchliffe {\it et al.}\cite{frank}, 
we adopt the same 
overall signal selection cuts which gave rise to only a small background 
contamination of mostly signal events: 
$\eslt >max(100\ {\rm GeV}, 0.2 M_{eff})$,
at least four jets with $E_T>50$ GeV, where the hardest jet has
$E_T>100$ GeV, transverse sphericity $S_T>0.2$ and $M_{eff}>800$ GeV.

In these events, we require at least two isolated leptons, and then plot 
the invariant mass of all same flavor/opposite sign dileptons. 
The results are shown in Fig. \ref{fig:mll}. In the case of the mSUGRA model,
frame {\it a}), 
there is a sharp peak at $m(\ell^+\ell^- )\sim M_Z$, which comes 
from $\tz_2\to \tz_1 Z^0$ decays where $\tz_2$ is produced in the gluino
and squark cascade decays. In the case of MWDM1 in frame {\it b}), 
we again see a $Z^0$ peak,
although here the $Z^0$s arise from $\tz_3$, $\tz_4$ and $\tw_2$ decays.
We also see the continuum distribution in 
$m(\ell^+\ell^- )<m_{\tz_2}-m_{\tz_1}=31.9$ GeV. The cross section 
plotted here is $\sim 0.05$ pb, which would correspond to 
5K events in 100 fb$^{-1}$ of integrated luminosity (the sample 
shown in the figure contains just 406 events). 
In frame {\it c})-- with a cross section of $\sim 0.05$ pb 
(but just 267 actual entries)-- we see again the $Z^0$ peak, 
but also we see again the 
$m(\ell^+\ell^- )<m_{\tz_2}-m_{\tz_1}=22.9$ GeV continuum.
In both these MWDM cases, the $m_{\tz_2}-m_{\tz_1}$ mass edge should 
be easily measurable. It should also be obvious that it is inconsistent 
with models based on gaugino mass unification, in that the
projected ratios $M_1:M_2:M_3$ will not be in the order $1:\sim 2:\sim 7$ as in
mSUGRA. Although the $\tz_2 -\tz_1$ mass edge will be directly measurable, 
the absolute neutralino and chargino masses will be difficult if not
impossible to extract at the LHC.
\FIGURE[htb]{
\epsfig{file=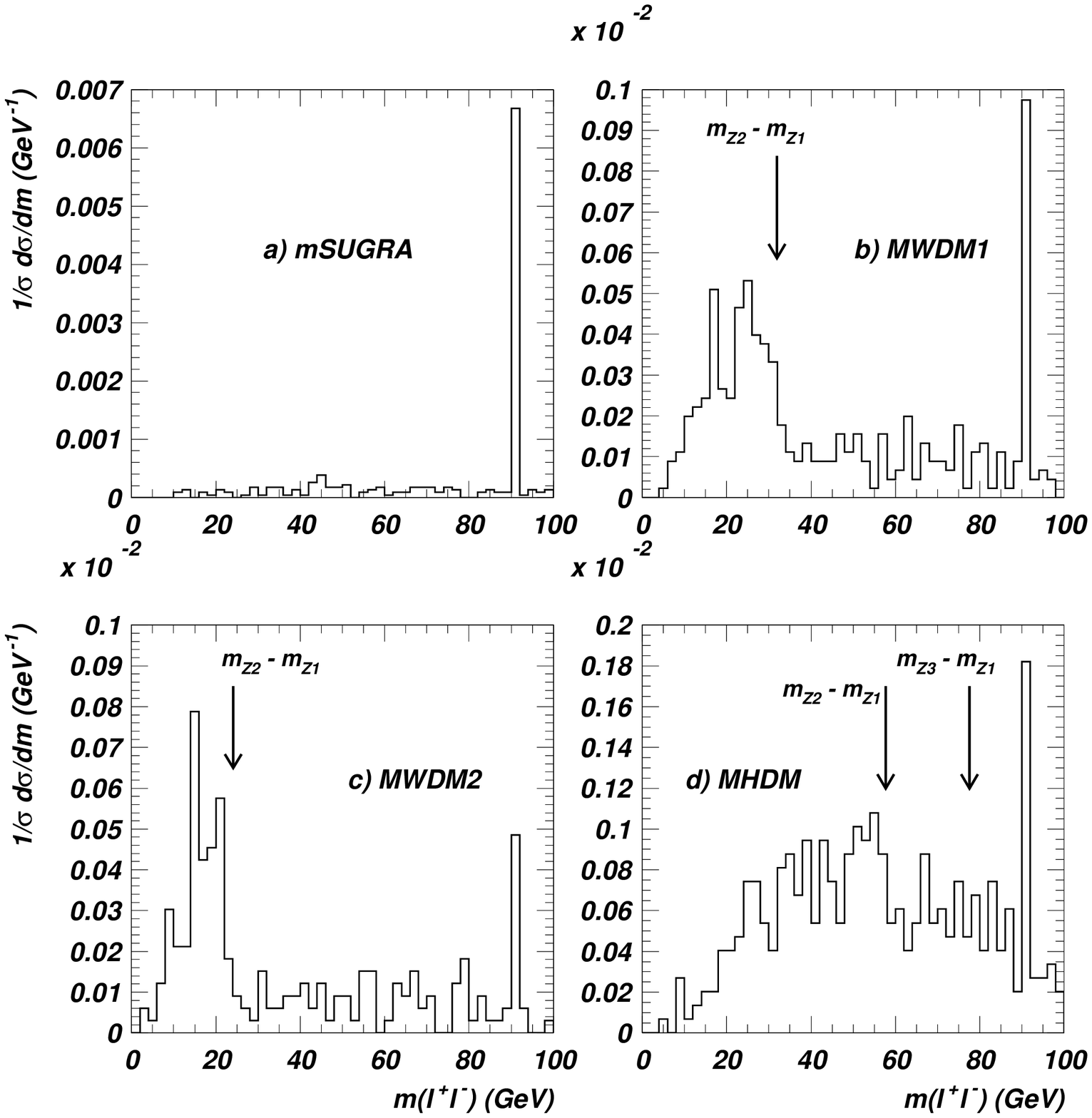,width=14cm} 
\caption{\label{fig:mll}
Distribution of same flavor/opposite sign dileptons from SUSY 
events at the CERN LHC 
from {\it a}) mSUGRA, {\it b}) MWDM1, {\it c}) MWDM2 
and {\it d}) MHDM cases 
as in Table \ref{tab:mwdm}.}}

In frame {\it d}), we show the spectrum from MHDM in the HB/FP region of 
the mSUGRA model. In this case, a $\tz_2 -\tz_1$ mass edge at 
57.7 GeV should be visible. It will be accompanied by other continuum
contributions, since in the case of MHDM with a small $\mu$
parameter, the $\tz_3$, $\tz_4$ and $\tw_2$ should all be relatively
light as well.

\subsection{Linear $e^+e^-$ collider}
\label{ssec:ilc}

At a $\sqrt{s}=500$ GeV ILC, the new physics reactions for the four
case studies shown in Table \ref{tab:mwdm} would 
include $Zh$, $\tw_1^+\tw_1^-$,
$\tz_1\tz_2$ and $\tz_2\tz_2$ production. It was shown in Ref. \cite{nlc}
that, in the case of a small $\tw_1 -\tz_1$ mass gap, 
chargino pair production events could still be identified 
above SM backgrounds.
The chargino and neutralino masses can be inferred from the resultant dijet
distribution in 
$\tw_1^+\tw_1^-\to (\bar{\ell}\nu_\ell\tz_1 )+(q\bar{q}\tz_1 )$
events\cite{jlc,bmt,nlc}. 
Alternatively, the chargino mass may be extracted 
from threshold cross section measurements when the 
CM energy of the accelerator is tuned to operate just above
$e^+e^-\to \tw_1^+\tw_1^-$ threshold.
These measurements should allow the absolute mass scale of the 
sparticles to be pinned down, and will complement the $\tz_2 -\tz_1$ 
mass gap measurement from the CERN LHC. 
The combination of $m_{\tz_2}$, $m_{\tw_1}$,
$m_{\tz_1}$ and $m_{\tz_2} -m_{\tz_1}$ measurements 
will point to whether or not gaugino
mass unification is realized in nature.

In addition, the $\tw_1^+\tw_1^-$, $\tz_1\tz_2$ and $\tz_2\tz_2$
production cross sections can all be measured as a function of beam
polarization at the ILC. In the mSUGRA model, since $\tw_1$ and $\tz_2$
are mainly wino-like, they will be produced at high rates for
left-polarized electron beams, but at low rates for right-polarized 
beams\cite{bmt}. 
The $\tz_1\tz_2$ production cross section also has a significant 
rise to it as
beam polarization parameter $P_L(e^-)$ is increased from -1 to +1.
These cross sections are plotted in frame {\it a}) of Fig. \ref{fig:ilc}.
In frame {\it b}), we show the same cross sections, except this time for
the case of MWDM1. The $\tw_1$ is still mainly wino-like, and so has a 
steeply rising cross section as $P_L(e^-)$ is increased. 
However, in this case
$\tz_1$ and $\tz_2$ both have non-negligible bino components, 
which enhances
their couplings to right-polarized electrons. 
Thus, $\sigma (e^+e^-\to \tz_1\tz_2)$ in the case of MWDM is a 
falling distribution $vs.\ P_L(e^-)$.
This is in fact borne 
out in frame {\it b}), and would be a strong signal for MWDM!
In frame {\it c}), we plot the corresponding cross sections for the case of
MWDM2. Again, $\tz_1\tz_2$ has a (slightly) 
falling cross section versus $P_L(e^-)$,
indicating once again the presence of MWDM.
In frame {\it d}), we show the corresponding cross sections for the case
of MHDM. In this case, numerous other reactions such as
$\tw_1^+\tw_2^-$, $\tz_1\tz_3$ and $\tz_2\tz_3$ should 
likely be kinematically accessible, and their presence will help
serve to distinguish MHDM from MWDM.
\FIGURE[htb]{
\epsfig{file=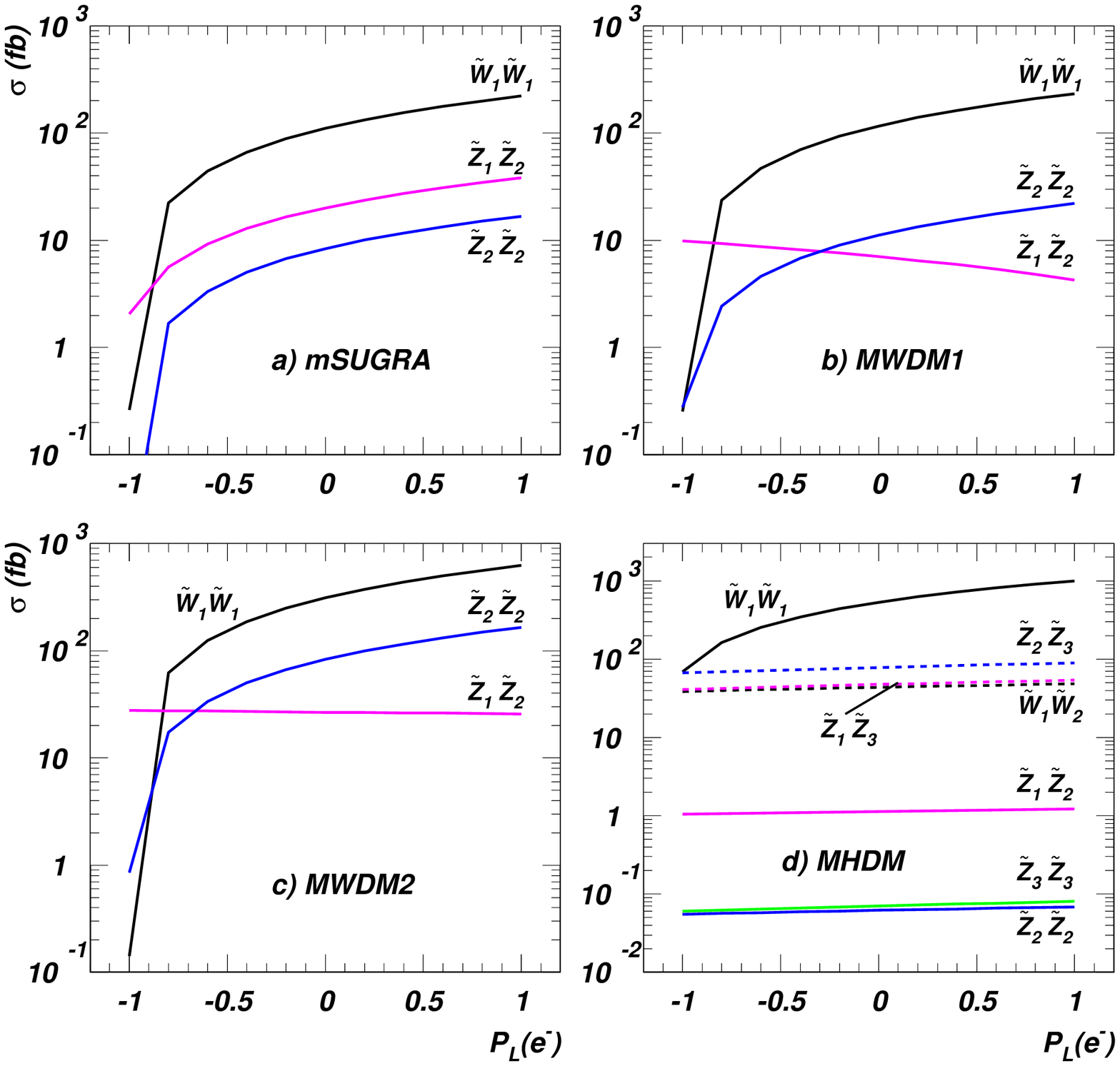,width=14cm} 
\caption{\label{fig:ilc}
Plot of cross section for $e^+e^-\to \tw_1^+\tw_1^-$, $\tz_1\tz_2$ and 
$\tz_2\tz_2$ versus electron beam polarization $P_L(e^-)$ for a 
$\sqrt{s}=500$ GeV ILC for {\it a}) mSUGRA, {\it b}) MWDM1,
{\it c}) MWDM2 and {\it d}) MHDM with parameters as in table \ref{tab:mwdm}.}}

While a combination of mass measurements at LHC and ILC would help to pin down
the properties of MWDM, it is worth considering whether the case of MWDM
can be confused with the case of MHDM, such as occurs in the HB/FP region
of the mSUGRA model, or in models with non-universal 
Higgs masses\cite{ellis,nuhm}.
To answer this, we plot in Fig. \ref{z2z1mw} the $\tz_2 -\tz_1$ mas gap
versus $m_{\tw_1}$ for MWDM scenarios which yield $\Omega_{\tz_1}h^2=0.11$,
against mSUGRA models in the HB/FP region which also give
$\Omega_{\tz_1}h^2=0.11$. We see that the MWDM points can span the entire
range of $m_{\tw_1}$ values shown, but that their $\tz_2 -\tz_1$ mass
gap is generally of order 15-40 GeV. Models with higher mass gaps are 
usually due to an overlap of MWDM with stau co-annihilation or $A$-funnel
annihilation. 
The general trend for $m_{\tz_2} -m_{\tz_1}$ in the 
MWDM scenario is dictated by the interplay of wino coannihilations and 
of the growing wino component, both functions of the mass gap, 
which suppress  $\Omega_{\tz_1}h^2$ to the required level. 
In contrast, the $\tz_2 -\tz_1$ mass gap associated
with MHDM in the HB/FP region is generally or order 40-80 GeV, 
at least until very large values of $m_{\tw_1}\agt 600$ GeV are 
generated. The largest mass gaps appear beyond the top quark mass 
threshold, whose effect is greatly enhanced, with respect to the 
MWDM case, due to $Z$ and Higgs $s$-channel exchanges. 
At larger neutralino masses, the $\tz_2 -\tz_1$ mass gap 
for MHDM shrinks to lower values, since a larger and larger 
higgsino fraction and stronger neutralino/chargino coannihilations 
are needed to fulfill the WMAP bound. 
Eventually, a pure higgsino LSP (with $m_{\tz_2} -m_{\tz_1}$ of the 
orders of few GeV) is needed to give $\Omega_{\tz_1}h^2=0.11$, 
at $m_{\tz_1}\sim 1$ TeV. For
$m_{\tw_1}\sim 600-800$ GeV, the MWDM and MHDM $\tz_2 -\tz_1$ mass gaps
overlap. In the large mass case, however, 
the two scenarios could still be differentiated
by the remaining sparticle mass spectrum ({\it e.g.} $\tz_3$ would be light
in the case of MHDM and heavy in the case of MWDM) and by the dependences 
of cross sections on electron beam polarization (if an energetic enough
$e^+e^-$ collider is built!). 
\FIGURE[htb]{
\epsfig{file=deltam_mw.eps,width=11cm} 
\caption{\label{z2z1mw}
Correlation between $m_{\tz_2}-m_{\tz_1}$ and $m_{\tw_1}$
in models with MWDM and MHDM in the HB/FP region.}}

\section{Conclusions}
\label{sec:conclude}

In this paper, we have considered the phenomenological 
consequences of mixed wino dark matter. MWDM occurs in models with gaugino
mass non-universality. MWDM may be obtained by modifying the 
paradigm mSUGRA model by either increasing the GUT scale value of $M_1$
or by decreasing $M_2$ until a sufficiently wino-like LSP is obtained
as to fulfill the WMAP measured value of $\Omega_{CDM}h^2\sim 0.11$.
If DM in nature is indeed composed of MWDM, then a number of consequences
occur. In the sparticle mass spectrum, 
the $\tz_2 -\tz_1$ and $\tw_1 -\tz_1$
mass gaps are expected to be reduced compared to what is 
expected in models 
with gaugino mass unification and a large $\mu$ parameter. Also, left- and
right- sleptons are expected to be more nearly mass degenerate.

If MWDM comprises the dark matter of the universe, 
then both direct and indirect dark matter detection rates are expected 
to be enhanced compared to expectations from the mSUGRA model.
However, to really pinpoint the existence of a partially wino-like
$\tz_1$, collider experiments will be needed. The CERN LHC should be able
to measure approximately the value of $m_{\tg}$, and in MWDM scenarios, 
also the $\tz_2 -\tz_1$ mass gap from the dilepton spectrum from
$\tz_2\to \ell\bar{\ell}\tz_1$ decay. These measurements should be enough 
to establish whether gaugino mass unification holds. Ultimately, a
linear $e^+e^-$ collider, the ILC,  
operating above $\tw_1^+\tw_1^-$ and $\tz_1\tz_2$
thresholds will be needed. The ILC should be able to measure the absolute
$\tw_1$, $\tz_1$ and $\tz_2$ masses. The dependence of the associated
production cross sections on the electron beam polarization
will point conclusively to the existence of MWDM.

\section*{Acknowledgments}
 
We thank J. O'Farrill and X. Tata for conversations.
This research was supported in part by the U.S. Department of Energy
under contract number DE-FG02-97ER41022.
	
%

\end{document}